\documentclass[eqsecnum,amsmath,preprintnumbers,superscriptaddress,nofootinbib,aps,twocolumn,10pt]
{revtex4} 
\pdfoutput=1
\usepackage{tikz}
\usetikzlibrary{calc,trees,positioning,arrows,chains,shapes.geometric,decorations.pathreplacing,decorations.pathmorphing,shapes,matrix,shapes.symbols}

\usepackage{bm}
\usepackage{MnSymbol}
\usepackage{amsmath}
\usepackage{graphicx}
\usepackage{feynmp}
\usepackage{hyperref}
\usepackage{bbold}
\usepackage{eufrak}
\usepackage{upgreek}
\usepackage{mathrsfs}
\usepackage{stmaryrd,scalerel}
\usepackage{mathtools}
\usepackage{enumitem}
\usepackage{color} 
\usepackage{etex}
\usepackage{morefloats}

 \usepackage{longtable}

\def\beq{\begin{equation}}
\def\eeq{\end{equation}}

\def\bea{\arraycolsep .1em \begin{eqnarray}}
\def\eea{\end{eqnarray}}

\def\tr{{\rm tr}}

\def\!!!{\stackrel{!}{=}}

\def\G{ \mathfrak{G}}

\def\a{\alpha}
\def\b{\beta}
\def\c{\gamma}

\newcommand{\g}{\mathfrak{g}}
\newcommand{\h}{\mathfrak{h}}

\def\de{\delta}

\def\eps{\epsilon}

\def\S{ \mathcal{S}}

\def\X{\mathcal{X}}
\def\I{ \mathfrak{E}}
\def\A{ {\rm a} }

\def\ha{ {\hat{\alpha}}}
\def\hb{ {\hat{\beta}}}
\def\hc{ {\hat{\gamma}}}

\def\im{ {\rm i }}

\def\nn{ \nonumber \\}

\def\eq#1{(\ref{#1})}

\def\s0#1#2{\mbox{\small{$ \frac{#1}{#2} $}}}
\def\0#1#2{\frac{#1}{#2}}

\def\grgl{\:\hbox to -0.2pt{\lower2.5pt\hbox{$\sim$}\hss}{\raise3pt\hbox{$>$}}\:}
\def\klgl{\:\hbox to -0.2pt{\lower2.5pt\hbox{$\sim$}\hss}{\raise3pt\hbox{$<$}}\:}

\begin{document}
\title{Gauge invariant effective actions for dressed fields}
\author{Kevin Falls}
\address{\footnotesize\mbox{Instituto de F\'isica, Facultad de Ingenier\'ia, Universidad de la Rep\'ublica, J.H.y Reissig 565, 11300 Montevideo, Uruguay}}

\date{\today}

\begin{abstract}
A new fundamental form of the path integral for theories with local symmetry is introduced. It is utilised to construct effective actions that generate correlation functions of dressed fields in Yang-Mills theories and quantum gravity. The construction entails a novel BRST symmetric gauge fixing which imposes that the on-shell correlation functions are those of gauge invariant fields. We demonstrate that the effective actions are gauge and diffeomorphism invariant respectively, with appropriate transformations for the Faddeev-Popov ghosts and  Nakanishi-Lautrup fields. As a consistency check, the on-shell one-loop effective actions are shown to take the expected gauge independent form. The effective action which satisfies Zinn-Justin's master equation is also  gauge invariant with the anti-fields transforming accordingly. While our choice of gauge will in general be non-linear and non-local, we argue that these gauges are in fact stable under renormalisation if one allows for general renormalised field variables.

\end{abstract}

\maketitle

\newpage
\tableofcontents

\newpage

\section{Introduction}

In gauge theories and general relativity, the fundamental fields 
that describe the underlying dynamics are not directly observable. This is 
because they transform non-trivially under  gauge 
transformations and diffeomorphisms, rendering them unobservable in their 
own right. For a classical theory this means that solutions to the equations of motion are 
underdetermined, even when ample initial conditions are provided. 
Specifically, the initial data alone cannot uniquely determine an 
individual solution; instead, what is fixed is an equivalence class of 
solutions that can be distinguished from one another by gauge 
transformations. To obtain fields which are in principle observable, at least for a classical theory,
one can ``dress'' the fundamental field variables to obtain composite  fields which are gauge invariant.

In the corresponding quantum theories one usually fixes the gauge to obtain a gauge fixed theory including new ghosts fields which interact with the original fields \cite{Faddeev:1967fc}. The gauge fixed theory is no longer gauge invariant but does possesses BRST symmetry \cite{Becchi:1975nq,Tyutin:1975qk}. This new symmetry encodes the fact that the physical theory is independent of the gauge choice. While correlation functions of the field variables depend on the gauge, correlation functions of gauge invariant operators do not. Thus for the quantum field theory gauge invariance is replaced by  the principle that observables are independent gauge condition.

When choosing a gauge for a specific calculation there some things that one can take into consideration.
For example, beyond perturbation theory one has to account for Gribov copies \cite{Gribov:1977wm} and thus the usual BRST invariant theory is not exact for typical gauge choices. One might therefore consider some gauges that do not have this problem. Unfortunately, for Yang-Mills theories,  such gauges are necessarily non-linear or discontinuous  \cite{Singer:1978dk}.  Putting this issue aside, in practice gauges are often chosen to simplify a  particular calculation for an observable of interest. Furthermore, one could look for an optimised choice of gauge which minimises errors \cite{Stevenson:1981vj} by studying the gauge dependence of observables.

As an alternative, to having to choose a concrete  gauge, one could imagine approximation schemes where the choice of gauge is seen to cancel out of the final result, order by order in the approximation.
In the context of asymptotically safe quantum gravity, such an approximation has been put forward in \cite{Falls:2024noj}.
With some care one may be able to perform calculations without ever fixing the gauge explicitly. Indeed, this can be realised manifestly using the exact renormalisation group \cite{Morris:1998kz,Arnone:2002cs,Rosten:2006qx,Falls:2020tmj}.

Since ultimately one is interested in computing gauge invariant operators we might ask if there is a more direct way to compute correlation functions of a complete set of gauge invariant fields.
One set of gauge invariant operators are the dressed, aka relational, fields which are obtained by a dressing process from the original field variables. The process requires a choice of dressing and results in a composite operator which is gauge invariant. For example, in general relativity the role of the dressing is played by four scalar fields constructed from the metric and/or matter fields \cite{Komar:1958ymq,Bergmann:1960wb,Bergmann:1961wa,Kuchar:1990vy,Brown:1994py, Rovelli:2001bz,Rovelli:1990ph,Rovelli:1990pi,Torre:1994ef,Torre:1993fq,    Dittrich:2004cb,Dittrich:2005kc,Westman:2007yx,Giddings:2005id,Hoehn:2020epv,Goeller:2022rsx}. Crucially they should behave such that they provide a physical coordinate system.
This being the case, each event in spacetime is then uniquely labeled by the values of the four scalars. 
The dressing process then consists of transforming the original field variables into the  physical coordinate system obtaining a composite field.
 An analogous dressing process can be performed for gauge theories where the role of the dressing is played by composite fields which are ``dynamical'' group elements. This generalises similar proposals for gauge invariant composite fields in gauge theories e.g.  \cite{Frohlich:1981yi,Zwanziger:1990tn,Lavelle:1995ty,Bagan:1999jk,Capri:2005dy,Capri:2015ixa,Capri:2016ovw,Dudal:2019pyg,Dudal:2020uwb,Dudal:2021pvw,Dudal:2021dec,Dudal:2023jsu,Maas:2016qpu,Maas:2017xzh,Maas:2020kda}.
 A recent discussion on the construction of gauge-invariant dressed operators in gauge theories and gravity has been given in \cite{Grassi:2024vkb}.

 In order to compute the correlation functions of dressed fields \cite{Brunetti:2013maa,Brunetti:2016hgw,Frob:2017lnt,Frob:2017gyj,Frob:2021ore,Frob:2021mpb,Frob:2023awn} one can first compute correlation functions of the gauge variant fields and then from these determine the correlation functions of the dressed fields. Since the latter are composite operators, this will ultimately involve both the renormalisation of the original generating functional of correlation functions and the further renormalisation process which deals with the composite fields.

The purpose of this paper is to write down a generating functional for the dressed correlation functions directly. 
The key idea is to use a specific gauge fixing condition that ensures that the fields coincide with the dressed fields. This leads naturally also to a different view point: we can view these gauges as providing the defining, although implicit, fundamental form of the path integral for any choice of complete gauge fixing. More specifically, if we insist that we have chosen a single representative, from each equivalence class of fields under gauge transformations, then we implicitly define a dressing and hence a dressed field. 
Therefore there is a correspondence between a choice of gauge and a choice of dressing.
  This is only formal in the sense that in practice one may want to explicitly choose the gauge or explicitly choose the dressing.

We have different motivations for our formalism. On the one hand, if one is interested in a particular dressed field, we can compute the correlation functions of that field directly within perturbation theory, for example. Another application is to keep the dressing unspecified and just use its properties (i.e. its transformation law), treating the choice as an unphysical gauge condition.  Observables do not depend on the gauge, so we can endeavour to compute observables without ever fixing the gauge at any stage.
Such observables include scattering amplitudes, quantum corrections to classical observables \cite{Frob:2021mpb}, as well as scaling dimensions of gauge invariant operators \cite{Baldazzi:2021fye}.       

 As we shall see, our effective actions enjoy gauge and diffeomorphism invariance respectively, in addition to the usual BRST symmetry.  Thus, we expect renormalised actions to maintain these powerful symmetries if a symmetric regulator is chosen. The price to pay is that the gauges are non-linear and thus we can expect that the renormalisation process is quite complex and we might expect that the simple tree-level form of the BRST invariant action will be deformed once quantum corrections are taken into account. On the other hand BRST exact terms  in the effective action are redundant; meaning they can be removed by a field redefinition.  This observation will lead us to the notation of ``quantum gauge invariance'' which we define as to mean that  independence is realised not in a trivial classical sense but with the general requirement that a field reparameterisation is needed. The idea that the gauge is redundant has been put forwarded in the context of  the essential renormalisation group \cite{Baldazzi:2021ydj}. Similar remarks have been made in \cite{Ihssen:2024ihp} and explored in detail recently \cite{Ihssen:2025cff}.

Our approach is quite orthogonal to the geometrical Vilkovisky-DeWitt approach \cite{Vilkovisky:1984st,DeWitt:2003pm,Lavrov:1988is,Buchbinder:1989ha,Odintsov:1989gz,Odintsov:1991yx,Kunstatter:1991kw,Burgess:1987zi,Pawlowski:2003sk,Branchina:2003ek,Donkin:2012ud}.
In particular we do not seek a gauge or parameterisation independent effective action. In the end all effective actions generate the correlation of some field variables and there is no ``unique'' choice for such a set of variables.
Our philosophy seeks an elegant  way to express the freedom to parameterise gauge theories and to exploit this freedom as much as possible.  

We also do not attempt to split the field into a ``physical part'' and a ``gauge part'' to achieve gauge invariance invariance as was the idea of \cite{Wetterich:2016ewc}. Instead we exploit a non-linear gauge fixing which achieves gauge invariance without the use of a background field \cite{Abbott:1981ke}. A gauge invariant gauge fixing seems like a contradiction it is not since our gauge fixing indeed breaks gauge invariance. However, the breaking terms can be absorbed into the source and those the gauge is broken by a non-vanishing source.

In section \ref{sec:GT} we discuss gauge transformations in Yang-Mills and gravity and fix our conventions.
Adopting a condensed notation, we will largely be able to treat gauge transformations in Yang-Mills and diffeomorphisms in gravity in a single frame work. In section~\ref{sec:DF} we discuss the construction and properties of dressed fields and three kinds dynamical gauge transformations (where by dynamical we mean that they depend on the field variables), which we call dressing fields, undressing fields and exchangers.  The relationships amongst each of these fields and to the gauge variant dressed fields is detailed.  In the end of the section we derive useful expressions which are obtained by taking functional derivatives both with respect to the fields and with respect to the gauge parameters. In particular we obtain the general expression for the functional derivative of the fields transformed by a dynamical gauge transformation. In section~\ref{sec:FI} we write down our fundamental form of the functional integral using dressing fields to write the gauge condition. The transformation law for the measure is derived by demanding that the path integral is independent of the gauge. 
The relation to the Faddeev-Popov form of the functional integral is discussed in section~\ref{sec:FP}.
Two examples of dressing fields are given in section~\ref{sec:examples} corresponding to Landau gauge in Yang-Mills and harmonic gauge in gravity.
In section~\ref{sec:YMaction} we construct a BRST invariant bare action for Yang-Mills based on our form of the path integral introducing ghosts and Nakanishi-Lautrup fields. An analogous BRST action is given for gravity in section~\ref{Sec:BRSTactionGravity}. In section~\ref{sec:Sym} we discuss the symmetries of the actions showing that, along side BRST symmetry, we have gauge and diffeomorphism invariance. In section~\ref{sec:EA} we consider the generating functionals of connected correlation functions and the effective action and show that when can obtain correlation functions of dressed fields from these functionals. We then show that the effective actions are gauge and diffeomorphism invariant respectively in section~\ref{sec:Inv}. The  tree-level propagators and on-shell one-loop effective action are studied in section~\ref{sec:1YM} for Yang-Mills and in  section~\ref{sec:1Grav} for gravity.
Section~\ref{sec:BRST_Consequences} recalls some known results relating to gauge independence of the functional integral and the master equation for the effective action and discusses the gauge invariance of the action with anti-fields. Section~\ref{sec:Gauge_Condition} demonstrates that BRST exact terms, i.e. those related to the choice of gauge, can be removed by a field reparameterisation and hence that the choice of gauge can be made stable under renormalisation.
We briefly discuss how different dressed fields must be related  in section~\ref{sec:renGI}. In particular we point out that  as the continuum limit is taken these relations consist of singular gauge transformations.
   In section~\ref{sec:Diss} we end the paper with a short discussion. In appendix~\ref{App_YM} we collect some result we need for Yang-Mills and in appendix~\ref{Measure_Trans} we give details relating to the transformation of the measure.

\section{Gauge transformations}
\label{sec:GT}
Here we will work in a general setup so we can treat both diffeomorphism invariance of general relativity and the gauge invariance of Yang-Mills.
This will also allow us to see how structures emerge in general, by only using that the gauge transformations form a group.
We denote the fields by $\phi^a$ where $a$ is a DeWitt index and so includes the  indices which label the components and the continuous spacetime coordinates $x^\mu$ in a single label. A repeated DeWitt index implies a sum over the indices and an integral over spacetime. Lower case latin letters from the start of the alphabet, $a$, $b,$ $c$ etc., are DeWitt indices for the fields.  As two examples we take the fields $\phi^a$ to be the gauge fields in Yang-Mills $\phi^a = \A^i_\mu(x)$ and the metric in gravity $\phi^a=g_{\mu\nu}(x)$. Therefore, in the case of Yang-Mills\footnote{Here and throughout $\int_x \equiv \int {\rm d}^4x$ denotes the integral over spacetime.} $ J_a \phi^a =\int_x  J^\mu_i(x) \A^i_\mu(x) $ and in the case of gravity $J_a  \phi^a = \int_x  J^{\mu\nu}(x) g_{\mu\nu}(x) $.  We can include matter fields in the components of $\phi^a$ also, however, for simplicity we treat all components $\phi^a$ as bosons.
Furthermore, we will treat gauge invariance and diffeomorphism invariance separately. It should be obvious that we can generalise to the case of the standard model plus gravity with appropriate sign changes for fermions and having both symmetries present. At times we use ``gauge transformations'' to also mean diffeomorphisms; it should be clear from the context when we are referring specifically to Yang-Mills and when we are being inclusive.

  Gauge transformations can be specified by a set of parameters $\mathfrak{g}^\a$ where, as the transformations depend on the point in spacetime, $\alpha$ is also a DeWitt index (as are the other greek letters from the start of the alphabet $\beta$, $\gamma$ etc.).  The field transformed by a gauge transformation with parameters $\mathfrak{g}^\a$ is denoted by $\g_*\phi^a$.  By $\g_*\h^\a$ we denote the new parameters which result after the operation of group multiplication of two elements with parameters $\mathfrak{g}^\a$ and $\mathfrak{h}^\a$.  The parameters of the inverse transformation are given by 
  \beq
  \hat{\g}^\a \equiv (\g^{-1})^\a
  \eeq
 As such  
 \beq
 \g_*\hat{\g}^\a = \mathfrak{1}^\a
 \eeq
  where $ \mathfrak{1}^\a$ are the values of the parameters for the identity group element such that $ \mathfrak{1}_* \phi^a = \phi^a$. 
Since the transformations form a group we have associativity 
\beq \label{associativity}
(\g_*\h)_*\phi^a = \g_*(\h_* \phi)\,,
\eeq 
so we can omit the brackets, i.e.
\beq
(\g_*\h)_*\phi^a \equiv  \g_*\h_*\phi^a   \equiv \g_*(\h_* \phi)\,.
\eeq 
We shall often refer to a set of gauge transformation parameters $\g^\a$ as simply a gauge transformation since the former specify the latter.  

Let's now see how things work for Yang-Mills and for gravity.
\subsection{Yang-Mills}
 For $SU(N)$ invariant Yang-Mills theories an element of the group can be represented by a matrix 
\beq \label{verde}
{\bf G}(\g) = e^{ {\rm i}  \g^i {\bf t}_i}
\eeq
which is parameterised by $ \g^i$ and we reserve $i$, $j$ $k$ etc. for the indices that label the generators ${\bf t}_i$.
We use bold face for the matrices which have the same dimensions as the group elements and use upper case latin letters $I$, $J$ $K$ etc. to label their components e.g. $({\bf G})_{IJ}$. 
 Since the group acts locally we have that $\g^\a = \g^i(x)$ are the parameters. So $\a$ stands for both the index $i$ and the continuous spacetime coordinates $x^{\mu}$. From \eq{verde} we note that, for Yang-Mills, the parameters of the identity are zero $\mathfrak{1}^\a = \mathfrak{1}^i(x)  =0$ and the parameters of the inverse are $\hat{\g}^i(x) = - \g^i(x)$.
We can normalise the generators of the group ${\bf t}_i$ by 
\beq
\tr \, {\bf t}_i {\bf t}_j = \frac{1}{2} \delta_{ij}\,.
\eeq
The indices $i$, $j$ and $k$ are raised and lowered using the Killing metric $\delta_{ij}$ which is simply the Kronecker delta (so we won't need to worry about the position of these indices or the positions of the indices $I$,$J$ and $K$ for that matter).
As standard we define the structure constants of the Lie algebra $f^{ijk}$ by 
\beq
[{\bf t}_i ,{\bf t}_j ] = {\rm i} f^{ijk} {\bf t}_k
\eeq
where it follows that the coefficients $ f^{ijk}$ are totally antisymmetric.
For $SU(N)$ theories we have that 
\beq
\tr\, {\bf t}_i = 0\,.
\eeq

The product $\g_*\h^\a$ is defined via
\beq
{\bf G}(\g) {\bf G}(\h) =  {\bf G}(\g_*\h) \,.
\eeq 
More explicitly this implies 
\beq
{\rm i} \g_*\h^i {\bf t}_i =  \log(e^{ {\rm i}  \g^i {\bf t}_i} e^{ {\rm i}  \h^i {\bf t}_i})  \,,
\eeq
which in turn implies
\beq
 \g_*\h^i = -2 {\rm i} \tr  {\bf t}_i  \log(e^{ {\rm i}  \g^j {\bf t}_j} e^{ {\rm i}  \h^k {\bf t}_k})  \,,
\eeq
which we can evaluate using the Baker-Campbell-Hausdorff formula.
In practice we will make use of the simpler identity
\bea \label{amarillo}
e^{ {\rm i}  \g^j {\bf t}_j}  {\bf t}_i  e^{- {\rm i}  \g^j {\bf t}_j} 
& =&  \left(e^F(\g)\right)^{ik}  {\bf t}_{k}\nn
& =&    {\bf t}_{k} \left(e^{-F(\g)}\right)^{ki}\,,
\eea
with
\beq
\left(F(\g)\right)^i\,_k := f^{ijk} \g^j\,,
\eeq
which we prove in  appendix~\ref{App_YM}  (note that although $F(\g)$ is a matrix we don't write it in bold since it does not have the dimensions of ${\bf G}$). Since $F(\g)$ is anti-symmetric 
\beq
O(\g) = e^{F(\g)}
\eeq
is an orthogonal matrix, which we used to reach the second line in \eq{amarillo}. 

The gauge transformation of the fields $\g_* \phi^a$ is given by the action of the group element ${\bf G}(\g) $ on the field. In the case that $\phi^a = \A^i_\mu(x)$, writing ${\bf a}_{\mu} = \A^i_\mu {\bf t}_i $,  we have that
\beq \label{boldA_transform}
\g_*{\bf a}_{\mu} = {\bf G}(\g) {\bf a}_{\mu}  {\bf G}^{-1}(\g) + {\rm i} {\bf G}(\g) \partial_{\mu}  {\bf G}^{-1}(\g)
\eeq
which is equivalent to 
\beq  \label{gauge_transform_A}
\g_*\A^k_\mu =   \A^i_\mu  \left(e^{F(\g)}\right)^{ik}  + \partial_\mu \g^i \left(e^{F(\g)}\right)^{ik} \,,
\eeq
as we show in Appendix~\ref{App:TransA}.
One can then check that the  associativity property \eq{associativity} holds.

\subsection{Gravity}
For diffeomorphisms $\g_*\phi^a$ denotes the {\it push forward} and the parameters are the functions $\g^\a=\g^\mu(x)$ which define the diffeomorphism as a map between manifolds. We recall that as a map $\g$ goes between two manifolds,  $\g: \mathcal{M}_x \to \mathcal{M}_{y}$, and the inverse as a map $\hat{\g}$ goes the other way, $\hat{\g}:  \mathcal{M}_{y} \to \mathcal{M}_{x}$. Then the push forward $\g_*$ takes a geometric object on $\mathcal{M}_x$, e.g. a scalar $\sigma(x)$,  and ``pushes it forward'' onto $\mathcal{M}_{y}$ to obtain a geometric object on $ \mathcal{M}_{y}$   e.g. $\g_* \sigma(y) = \sigma(\hat{\g}(y))$. For diffeomorphisms we can also define the pull back as the inverse $\g^* = \hat{\g}_*$ which takes a geometric object on $\mathcal{M}_y$ e.g.  a scalar $\tau(y)$ and ``pulls it back'' onto    $\mathcal{M}_x$ e.g. $\g^*\tau(x) = \tau(\g(x))$. Using push forwards as the gauge transformations rather than pull backs is simply a convention. 

We then define $\g_*\h^\a$ as the composition of the maps
\beq \label{rosa}
\h_*\g^\mu(x) = \h^\mu(\g(x)) \,,
\eeq
which satisfies the requirement of  associativity \eq{associativity}, e.g. for a scalar we require that 
$
(\h_* \g)_* \sigma(x) = \h_* (\g_* \sigma(x))
$
which holds 
since
$
(\h_* \g)_* \sigma(x) = \sigma(\hat{\g}(\hat{\h}(x)) 
$
and
$
 \h_* (\g_* \sigma(x)) =  \h_* (\sigma(\hat{\g}(x)) =   \sigma(\hat{\g}(\hat{\h}(x))) 
$.

 For diffeomorphisms the gauge parameters of the identity are $\mathfrak{1}^\mu(x) = x^\mu$ and the inverse $\hat{\g}^\mu(x)$ satisfies
\beq
\hat{\g}^\mu(\g(x)) = x^\mu\,, \,\,\,\,\,  \g^\mu(\hat{\g}(x)) = x^\mu\,.
\eeq

 Note that in addition to tensors on a manifold, which map from spacetime to some ``data'',  we can consider geometric objects, such as events or world lines, which take some data and map it into spacetime e.g. an event maps to a point with coordinates $x^\mu(\mathcal{P})$ where $\mathcal{P}$ is some data that uniquely fixes the coordinates $x^\mu(\mathcal{P})$ of the event i.e. $\mathcal{P}$ specifies the point physically. A push forward then acts on the coordinates of an  event as $\g_*x^\mu(\mathcal{P}) =  \g^{\mu}(x(\mathcal{P})) \equiv y^\mu(\mathcal{P})$. As such we see that  \eq{rosa} conforms to treating the gauge parameters as coordinates of events in terms of how other parameters acts on them via  $\h_*$. We note that if we would use the convention of pull backs as the gauge transformation then associativity would require us to define 
 \beq
\h^*\g^\mu(x) = \g^\mu(\h(x)) 
 \eeq
which conforms to treating each component $\g^\mu(x)$ as a scalar and is not as intuitive as \eq{rosa}, hence our choice of convention.
The price of our convention of using push forwards is a relative minus sign between small gauge transformations and Lie derivatives which we will see later.

If we take the field to be the metric $\phi^a= g_{\mu\nu}(x)$, then the push forward is given by
\beq \label{push_forward_g}
 \g_*g_{\mu\nu}(x) = \frac{\partial \hat{\g}^\rho(x)}{\partial x^\mu}  \frac{\partial \hat{\g}^\lambda(x)}{\partial x^\nu} g_{\rho \lambda}(\hat{\g}(x))\,,
\eeq
which is the general expression for a push forward of a rank $(0,2)$-tensor.

\subsection{General assumptions}
Generalising our two examples we will assume that the transformed $\g_*\phi^a$ is affine 
\beq
 \g_*\phi^a  =  T^a\,_b[\g] \phi^b + T^a[\g]\,.
\eeq 
as will be the case for standard parameterisations of the fields in Yang-Mills and gravity, exemplified by the gauge field and the metric. 
For our examples the coefficients $T^a\,_b[\g] $ and $T^a[\g]$ can be read off from the transformation laws \eq{gauge_transform_A}  and \eq{push_forward_g}.
 Viewed as a matrix $T^a\,_b[\hat{\g}]$ is the inverse of $T^a\,_b[\g]$, i.e. 
\beq \label{Tinverse1}
T^a\,_b[\hat{\g}] T^b\,_c[\g] = \delta^a_c\,, \,\,\,\,   T^a\,_b[\g] T^b\,_c[\hat{\g}] = \delta^a_c\,,
\eeq
where $\delta^a_c$ is a product of a Dirac delta and the identity on the field space,
while 
\beq \label{Tinverse2}
T^a[\hat{\g}] = - T^a\,_b[\hat{\g}] T^b[\g]\,.
\eeq
When $\phi^a = \A^i_\mu(x)$ we note that the inverse $T^a\,_b[\hat{\g}]$ is also transpose of  $T^a\,_b[\g]$ i.e. it is an orthogonal matrix.

\section{Dressed fields}
\label{sec:DF}
In gauge theories only gauge invariant composite operators are observable. 
Consequently, as we have said, the fields $\phi^a$ are not themselves observables. 
However, by adopting a ``physical'' gauge condition the fields will be equal to dressed fields $\hat{\phi}^a$ when the  condition applies. Moreover when expressed in terms of unconstrained fields $\phi^a$ the dressed fields  $\hat{\phi}^a[\phi]$ are gauge invariant functionals of the latter i.e. 
$
\hat{\phi}^a[\g_*\phi] = \hat{\phi}^a[\phi]
$.
 In this section we will develop the treatment of dressed fields within our formalism. It can then be applied to the construction of a functional integral, which in turn is the foundation of our effective actions.
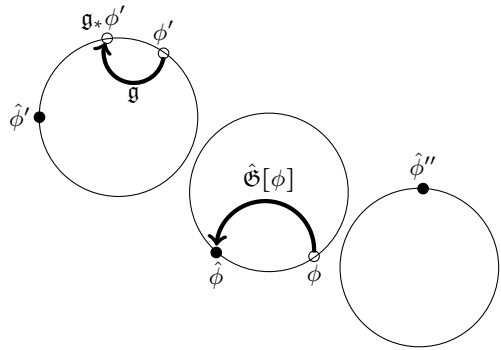
\begin{figure}
\begin{tikzpicture}
\draw (1,0) circle (30pt);
\draw[ultra thick, ->] (1.6,-0.8) arc (-3:175:0.65);
\draw[ultra thick, ->] (-0.4,1.80) arc (-8:-200:0.4);
\draw (-0.4,1.85) circle (2pt) node[anchor=south]{$\phi'$};
\draw (-1.15,2.05) circle (2pt) node[anchor=south]{$\g_*\phi'\,\,$};
\draw (-1,1) circle (30pt);
\draw (3,-1) circle (30pt);
\draw  (1.6,-0.85) circle (2pt) node[anchor=north]  {$\phi$} ;
\draw (1,0.20) node  {$\hat{\mathfrak{G}}[\phi]$} ;
\draw (-.8,1.3) node  {$\g$} ;
\filldraw[black] (-2.05,1) circle (2pt) node[anchor=east]{$\hat{\phi}'$};
\filldraw[black] (3.05,0.05) circle (2pt) node[anchor=south]{$\hat{\phi}''$};
\filldraw[black] (0.3,-0.8) circle (2pt) node [anchor=north]{$\hat{\phi}$};
\end{tikzpicture}
\caption{Schematic representation of configuration space. The circles are three gauge orbits with every pair of points the same circle being related by a gauge transformation, which we exemplify in the orbit on the left. Picking a section corresponds to choosing one point on each orbit to be the representative which are denoted by the solid points and labelled as $\hat{\phi}$,  $\hat{\phi}'$  and  $\hat{\phi}''$. In the middle orbit we show the action of the dressing field $\hat{\mathfrak{G}}[\phi]$ (portrayed by the bold arrow)  which takes a point $\phi$ (unfilled circle)  to its representative such that $\hat{\mathfrak{G}}[\phi]_* \phi = \hat{\phi}[\phi]$.  }
\label{gauge_orbits}
\end{figure}

There are two points of view we can adopt towards dressed fields. From a more technical point of view we can see them as being the outcome of a particular gauge fixing procedure. From a more physical perspective, we can think of them as relational or dressed observables that we can measure at least in the classical sense. From this point of view they represent a gauge invariant relation between a gauge variant field and a gauge variant dressing \cite{Rovelli:2013fga}. For the case of gravity the dressing is understood as a system of clocks and rods which form a physical coordinate system able to locate objects and define directions. Importantly, in a field theoretic formalism these coordinates are functionals of the fields themselves.  Transforming the metric into this coordinate system we get the dressed metric that is diffeomorphism invariant and hence physically well defined. The dressed metric therefore depends  on the fields that are used to construct the clocks and rods. These fields can be constructed from matter fields, from the metric itself, or from both. In gauge theories one can think of the dressing as a generalisation of a voltmeter that can measure a physical voltage difference taking into account all effects of the meter itself. Thus essentially the dressing is the ``gauge'' or measuring device that allows one to read off from its dial the ``value of gauge field'' in a gauge theory; hence the name ``gauge theory'' fits nicely when understood this way. 

\subsection{Representatives}
Let us start from the technical side. We consider the space of all field histories $\mathcal{H}_\phi$ which has $\phi^a$ as coordinates. Then we identify a choice of gauge with choice of a section of $\mathcal{H}_\phi$ when viewed as a fibre bundle. More specifically, we can break up the space $\mathcal{H}_\phi$ into equivalence classes of configurations under gauge transformations. Any two configurations $\phi$ and $\g_*\phi$ that are related by a gauge transformation a $\g$  are in the same equivalence class $\phi \sim \g_* \phi$.  The equivalence classes are the fibres of the fibre bundle, known as gauge orbits, and the base space is the space of orbits i.e. each equivalence class is a point in the base space. Picking a section means we pick one and only one history from each and every gauge orbit. A nice analogy is to think of each orbit as a constituency in a representative democracy and each representative being the elected member of parliament.   Each field configuration that lies in the section is the representative of the orbit to which it belongs. The choice of section is therefore analogous to the election of one member of parliament from every constituency who is doing the job of representing each of their constituents i.e. each $\phi^a$ such that $\phi^a \sim \hat{\phi}^a$.      For any choice of section a dressed field, understood as a functional $\hat{\phi}^a[\phi]$, is the map from the field configuration $\phi^a$ to its representative. Let us call $\hat{\mathfrak{G}}[\phi]$  the gauge transformation that takes $\phi^a$ to its representation then, by this definition,
\beq \label{dressing_field}
\hat{\phi}^a[\phi] = \hat{\mathfrak{G}}[\phi]_* \phi^a\,.
\eeq
A schematic of three gauge orbits with a choice of section is given in figure~\ref{gauge_orbits} with the orbits displayed as circles and black dots displaying the position of the representative in each of the orbits. The placement of the representatives is arbitrary from the technical point of view. 
Therefore as a purely technical device the function $\hat{\mathfrak{G}}[\phi]$ need not be a nice smooth function since any choice of a section is permissible. However, we will need that $\hat{\mathfrak{G}}[\phi]$ are suitably differentiable. Moving within an orbit, we have that
\beq \label{Trans_dressing_field}
 \hat{\mathfrak{G}}[\g_*\phi]=  \hat{\mathfrak{G}}[\phi]_* \hat{\g}\,,
\eeq
which again follows from the definition that we pick a section. A graphical proof of this law is given in figure~\ref{conejo}.
It also follows, as was mentioned, that the dressed field $\hat{\phi}^a[\phi]$ is a gauge invariant functional of the field $\phi^a$ which we can show in a few steps
\bea
\hat{\phi}^a[\g_*\phi]&=&  \hat{\mathfrak{G}}[\g_*\phi]_* \g_* \phi^a \nn
& =&  \hat{\mathfrak{G}}[\phi]_* \hat{\g}_* \g_* \phi^a \nn
&=&  \hat{\mathfrak{G}}[\phi]_* \phi^a\nn
  &=& \hat{\phi}^a[\phi]\,.
\eea
Indeed this is just the statement that each field in the same equivalence class has the same representative. 

\begin{figure}
\begin{tikzpicture}
\draw (1,0) circle (30pt);
\draw[ultra thick, ->] (1.6,-0.8) arc (-3:175:0.65);
\draw[ultra thick, ->] (2.1,-0.15) arc (30:-99:0.45);
\draw[ultra thick,gray, ->] (2.1,-.13) arc (45:-180:1.05);
\draw  (2.1,-0.85)  node {$\hat{\g}$} ;
\draw  (1.6,-0.85) circle (2pt) node[anchor=north]  {$\, \phi$} ;
\draw  (2.05,-0.1) circle (2pt) node[anchor=west]  {$\,\g_* \phi$} ;
\draw (1,0.20) node  {$\hat{\mathfrak{G}}[\phi]$} ;
\draw (1.5,-2.20) node  {$\hat{\mathfrak{G}}[\g_*\phi] $} ;
\filldraw[black] (0.3,-0.8) circle (2pt) node [anchor=east]{$\hat{\phi}\,$};
\end{tikzpicture}
\caption{A graphical proof of the transformation law for the dressing field $\hat{\mathfrak{G}}[\g_*\phi] = \hat{\mathfrak{G}}[\phi]_*\hat{\g}$: we can apply first $\hat{\g}$ to $ \g_*\phi$ and then apply $\hat{\mathfrak{G}}[\phi]$, as on the rhs, corresponding to following the route of the two black transformations. Alternatively we can apply $\hat{\mathfrak{G}}[\g_*\phi]$ to $\g_*\phi$, as on the lhs, corresponding to the single transformation in grey. }
\label{conejo}
\end{figure}
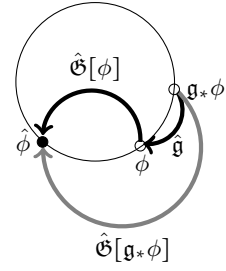

\begin{figure}
\begin{tikzpicture}
\draw (1,0) circle (30pt);
\draw[ultra thick, ->] (1.6,-0.8) arc (-3:175:0.65);
\draw[ultra thick, <-] (1.55,-0.95) arc (-18:-165:0.65);
\draw  (1.6,-0.85) circle (2pt) node[anchor=west]  {$\, \phi$} ;
\draw (1,0.20) node  {$\hat{\mathfrak{G}}[\phi]$} ;
\draw (1,-1.65) node  {$\mathfrak{G}[\phi]$} ;
\filldraw[black] (0.3,-0.8) circle (2pt) node [anchor=east]{$\hat{\phi}\,$};
\end{tikzpicture}
\caption{Schematic representation of the action of the dressing and undressing fields $\hat{\mathfrak{G}}[\phi]$ and $\mathfrak{G}[\phi]$ respectively expressed in equations \eq{dressing_field} and \eq{undressing_field}.   }
\label{undressing_gauge_orbit}
\end{figure}
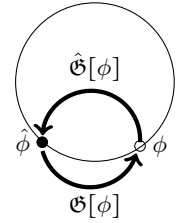

\subsection{Dressing and undressing fields}
\begin{figure}
\begin{tikzpicture}
\draw (1,0) circle (30pt);
\draw[ultra thick, <-] (1.6,-0.9) arc (-3:-180:0.65);
\draw[ultra thick, <-] (2.1,-0.15) arc (30:-99:0.45);
\draw[ultra thick,gray, <-] (2,-.1) arc (45:178:1.);
\draw  (2.1,-0.85)  node {$\g$} ;
\draw  (1.6,-0.85) circle (2pt) node[anchor=north]  {$\,\,\, \,\, \, \phi$} ;
\draw  (2.05,-0.1) circle (2pt) node[anchor=west]  {$\,\,\, \g_* \phi$} ;
\draw (1,-1.75) node  {$\mathfrak{G}[\phi]$} ;
\draw (1.1,0.40) node  {$\mathfrak{G}[\g_*\phi] $} ;
\filldraw[black] (0.3,-0.8) circle (2pt) node [anchor=east]{$\hat{\phi}\,$};
\end{tikzpicture}
\caption{A graphical proof of the transformation law for the undressing field $\mathfrak{G}^\alpha[\g_* \phi] = \g_*\mathfrak{G}^\alpha[ \phi] $: we can apply first $\mathfrak{G}^\alpha[ \phi] $ to $\hat{\phi}$ and then apply $\g$, as on the rhs, corresponding to following the route of the two black transformations. Alternatively we can apply $\mathfrak{G}[\g_*\phi]$ to $\hat{\phi}$, as on the lhs, corresponding to the single transformation in grey. }
\label{gato}
\end{figure}
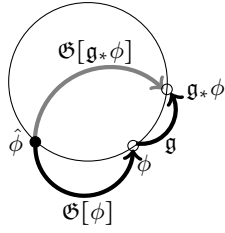

The more physical point of view is to think of $\hat{\mathfrak{G}}^\alpha[\phi]$ as fields that dress the gauge variant fields $\phi^a$ to form dressed fields. Thus we call composite fields  $\hat{\mathfrak{G}}^\alpha[\phi]$ with the transformation law \eq{Trans_dressing_field} {\it dressing fields}. The inverses  $\mathfrak{G}^\alpha[\phi]$ are called the {\it undressing fields} since they transform the dressed field back to the gauge variant field
\beq \label{undressing_field}
\mathfrak{G}[\phi]_* \hat{\phi}^a[\phi] = \phi^a\,.
\eeq
 Figure~\ref{undressing_gauge_orbit} shows a schematic representations of \eq{dressing_field} and \eq{undressing_field}.
By definition the undressing field $\mathfrak{G}^\alpha[\phi]$ transforms as 
\beq \label{Trans_undressing_field}
\mathfrak{G}^\alpha[\g_* \phi] = \g_*\mathfrak{G}^\alpha[ \phi] \,.
\eeq
A graphical proof of \eq{Trans_undressing_field} is given in figure~\ref{gato}.

From the physical point of view a dressing field is chosen explicitly as the first step to find the form of the dressed fields. In the more abstract point of view of  an arbitrary section is chosen and the dressing field is chosen implicitly. One might imagine that given a particular observable one is trying to compute that there is an optimal choice of dressing field which reduces the complexity of the calculation.  

\subsection{Exchangers}
\label{sec:Ex}

Dressing and undressing fields are types of dynamical gauge transformation, where the transformation parameters depend on the fields, which are distinguished by their respective transformation laws \eq{Trans_dressing_field} and \eq{Trans_undressing_field}. 
We define an {\it exchanger} as dynamical gauge transformation  $\mathfrak{E}[\phi]$ which transforms as\footnote{For gravity such dynamical diffeomorphisms have been introduced in \cite{Ferrero:2020jts}.} 
\beq \label{Trans_exchanger}
\mathfrak{E}[\g_*\phi] =\g_* \mathfrak{E}[\phi]_* \hat{\g}\,.
\eeq
Note that the inverse of a exchanger $\hat{\mathfrak{E}}[\phi]$ is also an exchanger i.e.
\beq\label{Trans_exchanger_hat}
\hat{\mathfrak{E}}[\g_*\phi] =\g_* \hat{\mathfrak{E}}[\phi]_* \hat{\g}\,.
\eeq
Exchangers can be understood as covariant gauge transformations since they transform the field into a composite operator  $\mathfrak{E}[\phi]_*\phi$  that transforms as the field does
\beq
 \mathfrak{E}[\g_*\phi] _* \g_*\phi = \g_*  \mathfrak{E}[\phi]_* \phi
\eeq

Let us explain why we call them exchangers. First we note that they provide a diffeomorphism of $\mathcal{H}_\phi$, i.e. a field reparameterisation, defined by
\beq \label{E_diiffeo}
\phi^a \to E^a[\phi]\equiv  \hat{\mathfrak{E}}[\phi]_*\phi^a
\eeq
 This diffeomorphism of $\mathcal{H}_\phi$ maps points in the same equivalence class to each other. Furthermore this map commutes with gauge transformations i.e.
 $E^a[\g_* \phi] = \g_* E^a[\phi]$ such that the same transformation $\g_*$ relates configurations before and after applying the field reparameterisation.
 In particular the diffeomorphism \eq{E_diiffeo} maps each representative $\hat{\phi}^a$ to another configuration in the same equivalence class and hence performs the analogy of a general election. Therefore
 \beq
 \hat{\varphi}^a = E^a[\hat{\phi}] = \hat{\mathfrak{E}}[\hat{\phi}]_*\hat{\phi}^a 
 \eeq
 is a new dressed field $\hat{\varphi}^a =  \hat{\varphi}^a[\phi]$ given by
 \beq
  \hat{\varphi}^a[\phi] =  \hat{\mathfrak{G}}[\phi]_*\hat{\mathfrak{E}}[\phi]_* \phi
 \eeq
 which we can write in terms of a new dressing field
 \beq
   \hat{\varphi}^a[\phi]  = \hat{\mathfrak{H}}[\phi]_*\phi
 \eeq
given by 
 \beq \label{H_transform}
  \hat{\mathfrak{H}}[\phi] =   \hat{\mathfrak{G}}^\alpha[\mathfrak{E}[\phi]_*\phi] =  \hat{\mathfrak{G}}[\phi]_*\hat{\mathfrak{E}}[\phi] 
 \eeq
 where we have used \eq{Trans_dressing_field} to reach the second equality.
Indeed using both \eq{Trans_exchanger_hat} and  \eq{Trans_dressing_field} we have
\bea
\hat{\mathfrak{H}}^\alpha[\g_*\phi]     &=& \hat{\mathfrak{G}}^\alpha[\g_* \phi]_* \hat{\mathfrak{E}}[\g_* \phi] \nn
&=&  \hat{\mathfrak{G}}^\alpha[\phi]_* \hat{\g}  _*\g_* \hat{\mathfrak{E}}[ \phi]_* \hat{\g}\nn
&=&\hat{\mathfrak{G}}^\alpha[\phi]_* \hat{\mathfrak{E}}[ \phi]_* \hat{\g}\nn
&=&  \hat{\mathfrak{H}}^\alpha[\phi]_*\hat{\g} 
\eea
which is the transformation law for a dressing field.
Moreover, we can construct all exchangers out of a dressing and undressing field by 
\beq 
   \mathfrak{E}[\phi] =  \mathfrak{H}[\phi]_*\hat{\mathfrak{G}}^\alpha[\phi]\,.
\eeq

In summary, given one choice of a section, corresponding to some choice of dressing field, we can use the exchanger to exchange the dressing field for another and hence change the choice of section i.e. choose a different set of representatives.  Thus exchangers are isomorphic to changing our choice of gauge and hence will play a role in determining the conditions for gauge independence in the next section. For a related discussion see \cite{Goeller:2022rsx} where a change in the dressed field by an exchanger i.e. $\hat{\phi} \to \mathfrak{E}[\hat{\phi}]\hat{\phi}$ is called a  change of dynamical reference frame.

\subsection{Dynamical gauge transformations in Yang-Mills theories}
Returning to our examples, for Yang-Mills theories $\hat{\mathfrak{G}}$ do not take values in the gauge group in its matrix representation. 
We can consider the {\it group dressing fields}  
\beq
{\bf D}[\phi] = {\bf G}_{\hat{\mathfrak{G}}[\phi]}
\eeq
  which transform as
\beq
{\bf D}[\g_*\phi]  = {\bf D}[\phi]  {\bf G}_{\hat{\mathfrak{g}}}\,.
\eeq
So, while the dressing fields $\hat{\mathfrak{G}}[\phi]$ transforms in a non-linear fashion in Yang-Mills, the group dressing field ${\bf D}[\phi]$ transforms linearly. Later we will use a DeWitt notation where the components of the group dressing fields are denoted by $D^{\ha}[\phi]$. In practice we will use the group dressing fields to construct our gauge fixing action.

\subsection{Dynamical gauge transformations in gravity}
 For diffeomorphisms the dressing fields transform as a collection of four scalars $\hat{\mathfrak{G}}^\a = \hat{\mathfrak{G}}^{\hat{\mu}}(x)$  such that 
\beq
\hat{\mathfrak{G}}^{\hat{\mu}}(x)|_{\phi \to \g_*\phi} = \hat{\mathfrak{G}}^{\hat{\mu}}(\hat{\g}(x))\,.
\eeq
The $\hat{\mu}$ index is therefore not a spacetime index but an index that labels four scalars.
 The undressing fields $\mathfrak{G}^\a = \mathfrak{G}^{\mu}(\hat{x})$  transform as the coordinates  of ``events''  
\beq
\mathfrak{G}^{\mu}(\hat{x})|_{\phi \to \g_*\phi}  = \g^\mu(\mathfrak{G}^{\mu}(\hat{x}))\,.
\eeq
In particular $\mathfrak{G}^{\mu}(\hat{x})$ are the coordinates of the points where the four scalars take the values $\hat{x}^{\hat{\mu}}$.
The exchangers transform as dynamical diffeomorphisms \cite{Ferrero:2020jts}   
\beq
\hat{\mathfrak{E}}^{\mu}(x)|_{\phi \to \g_*\phi}  =   \g^\mu(\hat{\mathfrak{E}}(\hat{\g}(x)))\,,
\eeq
with left and right composition. 

In order to have a common notation with Yang-Mills we will write the dressing fields in gravity as $\hat{\mathfrak{G}}^\a = D^{\hat{\mu}}(x)$
and also use a hatted index in the DeWitt notation so that
\beq
\hat{\mathfrak{G}}^{\hat{\a}}[\phi] = D^{\ha}[\phi] = D^{\hat{\mu}}(x)[\phi]\,.
\eeq
 This introduces some redundancy in our notation since we have two symbols for dressing fields in gravity. Roughly speaking we use $\hat{\mathfrak{G}}^{\hat{\a}}[\phi]$ to stress that it is a dynamical gauge transformation while developing the formalism  and $D^{\ha}[\phi]$ in the context of the BRST invariant action, which is used in practice and where the role is the same as ${\bf D}[\phi]$ in Yang-Mills.

\subsection{Functional derivatives}

One purpose treating gauge transformations in terms of the parameters, which are themselves composite fields, is that we can take functional derivatives with respect to the parameters. Hence we have that 
\beq
\frac{\delta \g^\a}{\delta \g^\b} = \delta^\a_\b\,,
\eeq
where for Yang-Mills
\beq
 \delta^\a_\b = \delta(x-y) \delta^i_j\,,
\eeq
with $\delta(x-y)$ and $\delta^i_j$ denoting the Dirac delta and Kronecker delta respectively.
Similarly, for gravity
\beq
 \delta^\a_\b = \delta(x-y) \delta^\mu_\nu\,.
\eeq
Now we can take functional derivatives with respect to the fields $\phi^a$ and with respect to the parameters $\g^\a$.
Moreover the parameters can depend on the fields when they are dressing fields, undressing fields or exchangers.
Here we will derive some useful results which can be used in these cases.

Let's define
\beq
 T^a_\a[\g,\phi]:=
\frac{\delta (\mathfrak{g}_*\phi^a)}{\delta \g^\a} 
\eeq
and
\beq \label{Tphi_def}
T^a_\a[\phi] :=  T^a_\a[\mathfrak{1},\phi]
\eeq
For gravity, when the field is the metric,
\beq 
T^a_\alpha[\phi] \eps^\a = - \mathcal{L}_{\eps} g_{\mu\nu} = -\nabla^{(g)}_\mu \eps_\nu - \nabla^{(g)}_\nu \eps_\mu 
\eeq
is minus the Lie derivative of the metric with $\nabla^{(g)}_\mu$ denoting the covariant derivative compatible with the metric (the minus is due to the fact that we have chosen to use the push forward as the gauge transformation rather than the pull back).
While for Yang-Mills we have that
\beq
T^a_\alpha[\phi]  \eps^\a = \partial_\mu \eps^i + f^{ijk}\A^j_\mu \eps^k  \equiv \nabla^{(A)}_\mu \eps^i  \,,
\eeq
with $\nabla^{(\A)}_\mu$ denoting the gauge covariant derivative.
Now consider
\beq
\h_*\phi^a= \h_* \hat{\g}_*\g_* \phi^a
\eeq
taking a functional derivative with respect to $\h$ and then setting $\h = \g$ we find that 
\beq
 T^a_\a[\g,\phi]  = U_\a\,^\b[\hat{\g}]   T_\b^a[\g_*\phi]
 \eeq
 where
 \beq
  U_\c\,^\a[\hat{\g}]  :=   \left. \frac{\delta  \h_*\hat{\g}^\a }{\delta  \h^\c}\right|_{\h = \g }
 \eeq
 and therefore 
\beq
 T^a_\a[\g,\hat{\g}_*\phi] = U_\a\,^\b[\hat{\g}] T_\b^a[\phi]
\eeq  
 Next let's consider  the identity 
 \beq
 \g[\phi]_*\hat{\g}[\phi]_* \phi^a =\phi^a
 \eeq
 where we let the $\g[\phi]$ be arbitrary functionals of the fields.
 Taking a derivative of the above identity with respect to the fields we find that
 \beq
   T^a\,_c[\g]  \frac{ \delta \hat{\g}_* \phi^c}{\delta \phi^b} +  \frac{\delta \g^\a}{\delta \phi^b} U_\a\,^\b[\hat{\g}] T_\b^a[\phi] = \delta^a_b\,,
 \eeq
which can be rewritten as
\beq
\frac{ \delta \hat{\g}_* \phi^a}{\delta \phi^b}  =    T^a\,_c[\hat{\g}] \left(\delta^c_b  -  \frac{\delta \g^\a}{\delta \phi^b} U_\a\,^\b[\hat{\g}] T_\b^c[\phi]  \right)\,,
\eeq
equally, swapping $\g$ with $\hat{\g}$ we have
\beq \label{Azul}
\frac{ \delta \g_* \phi^a}{\delta \phi^b}  =    T^a\,_c[\g] \left(\delta^c_b  -  \frac{\delta \hat{\g}^\a}{\delta \phi^b} U_\a\,^\b[\g] T_\b^c[\phi]  \right)\,.
\eeq
\section{Functional integral}
\label{sec:FI}
In this section we will construct the functional integral putting aside the issue of renormalisation.  
Since each choice of section is just a choice of the a representative for each equivalence class, we are at liberty to choose one without any physical effects.
We can define the gauge independent functional integral by 
\beq \label{Z_can}
\mathcal{Z} = \int d \phi \, \tilde{\mu}[\phi]\, \delta(\mathfrak{1} - \hat{\mathfrak{G}}[\phi]) e^{-S[\phi]}
\eeq
which restricts the integral to a single section. 
Here $S[\phi]$ is a gauge invariant action and  $\tilde{\mu}[\phi]$ is a measure factor, whose properties we will determine below. These properties ensure that $\mathcal{Z}$ is not dependent on the gauge i.e. on the choice of section.  
Let's note that our form of the functional integral can be seen as a specific class of gauge fixed functional integrals without the need for a FP ghost determinant. The FP ghost operator is given by 
\bea
Q^{\a}\,_\b[\phi] &:=& \left. - \frac{\delta}{\delta \g^\b} (\hat{\mathfrak{G}}[\phi]_* \hat{\g})^{\a} \right|_{\g= \mathfrak{1}}\\
&=& \left.  \frac{\delta}{\delta \g^\b} (\hat{\mathfrak{G}}[\phi]_* \g)^{\a} \right|_{\g= \mathfrak{1}}
\eea
however when $\hat{\mathfrak{G}}[\phi] = \mathfrak{1}$ then $Q^{\a}\,_\b[\phi] = \delta^\a_\b$ is the identity matrix.  Hence 
$  \delta(\mathfrak{1} - \hat{\mathfrak{G}}[\phi])  \det Q^\a\,_\b =  \delta(\mathfrak{1} - \hat{\mathfrak{G}}[\phi])$.

As we have said, the measure factor $\tilde{\mu}[\phi]$ must transform such that $\mathcal{Z}$ is independent of the choice of section i.e. independent of the dressing field $\hat{\mathfrak{G}}[\phi]$. Finding this transformation law is a key result of this paper.
It guarantees that we can equally write \eq{Z_can} as
\beq \label{Z_canH}
\mathcal{Z} = \int d \phi \, \tilde{\mu}[\phi]\, \delta(\mathfrak{1} - \hat{\mathfrak{H}}[\phi]) e^{-S[\phi]}
\eeq
for any dressing field $\hat{\mathfrak{H}}[\phi]$ without using BRST invariance. 
By making a change of variables in the functional integral $\phi \to \mathfrak{E}[\phi]_*\phi$, we see that this will be the case if, for the exchanger $\mathfrak{E}[\phi] = \mathfrak{H}[\phi]_*\hat{\mathfrak{G}}^\alpha[\phi]$, it holds that 
\beq \label{Rojo}
 d \left(\mathfrak{E}[\phi]_*\phi \right) \, \tilde{\mu}[\mathfrak{E}[\phi]_*\phi] = d \phi \, \tilde{\mu}[\phi] \,.
\eeq
To see this one uses the gauge invariance of $S[\phi]$ and \eq{H_transform} which implies  $ \delta(\mathfrak{1} - \hat{\G}[ \mathfrak{E}[\phi]_*\phi])=  \delta(\mathfrak{1} - \hat{\mathfrak{H}}[\phi])$.
From \eq{Azul} we have that
\bea
 d (\mathfrak{E}[\phi]_*\phi)  &=& d\phi \det T^a\,_b[\mathfrak{E}] \\
 && \times \det\left( \delta^a_b  -  \frac{\delta \hat{\mathfrak{E}}^\a}{\delta \phi^b} U_\a\,^\b[\mathfrak{E}] T_\b^a[\phi] \right) \nonumber
\eea
Then a short calculation, which we carry out in Appendix~\ref{Measure_Trans}, reveals that 
\beq \label{Rosa}
 d (\mathfrak{E}[\phi]_*\phi)  =d\phi \det T^a\,_b[\mathfrak{E}] \det T^\a\,_\b[\hat{\mathfrak{E}}] \,,
\eeq
where we define
\beq
T^\a\,_\b[\g]:=  \left.  \frac{\delta \g_*\h_*\hat{\g}^\a}{ \delta \h^\b} \right|_{\h= \mathfrak{1}}\,.
\eeq
Noting that, by the chain rule, $T^\a\,_\b[\g]$ satisfies
\beq
T^\a\,_\b[\g] T^\b\,_\c[\hat{\g}] = \delta^\a_\c\,,
\eeq
let's then introduce symmetric invertible two point functions $\gamma_{ab}[\phi]$ and $\eta_{\a\b}[\phi]$ which transform as
\beq
\gamma_{ab}[\phi] = \gamma_{ab}[\g_*\phi] = T^c\,_a[\hat{\g}] \gamma_{cd}[\phi] T^{d}\,_b[\hat{\g}] \,,
\eeq 
and
\beq
\eta_{\a\b}[\g_*\phi] = T^\c\,_\a[\hat{\g}] \eta_{\c\de}[\phi] T^{\de}\,_\b[\hat{\g}] \,.
\eeq
Then by setting \cite{Falls:2020tmj}
\beq
\tilde{\mu}[\phi] = \frac{\sqrt{\det \gamma_{ab}[\phi]}}{\sqrt{\det \eta_{\a\b}[\phi]}}
\eeq
we satisfy \eq{Rojo} and hence $\mathcal{Z}$ does not depend on the choice of section as required.

For Yang-Mills we have that 
\beq \label{Talphabeta_YM}
T^{\a}\,_\b v^\b= \left(e^{-F(\g(x))}\right)^{ij}v^j(x)
\eeq
as we demonstrate in Appendix~\ref{App_YM}.
For gravity
\beq
T^{\a}\,_\b v^\b = \frac{\partial \g^\mu(x)}{\partial x^\nu}  v^{\nu}(\hat{\g}(x))
\eeq
which we recognise as the push forward of a contra-variant vector.

It is worth noting that for gravity the measure typically has to be non-trivial.
The metrics $\gamma_{ab}$ and $\eta_{\a\b}$ can always be chosen to have a minimal form for which it agrees with 
the measure of Fujikawa \cite{Fujikawa:1983im,Fujikawa:2004cx}.  An in depth discussion of the path integral in quantum gravity is given in \cite{Bonanno:2025xdg}.

 while for Yang-Mills one can satisfy \eq{Rojo} simply by setting $\eta_{\a\b} \propto \delta_{\a\b}$ and $\gamma_{ab} \propto  \delta_{ab}$ with field independent coefficients and hence $\tilde{\mu}$ is a constant. This follows since for Yang-Mills the transposes of $T^{\a}\,_\b$ and $T^a\,_b$  are  their respective inverses.
Nonetheless for the purposes of regularisation choosing $\gamma_{ab}$ and $\eta_{\a\b}$ to be non-trivial can be desirable \cite{Falls:2020tmj}. 

Here, consider the unregulated forms for Yang-Mills  
\beq
\gamma^{\mu,\nu}_{i,j}(x,y) = \frac{\mu^2}{{\rm g}} \delta^{\mu \nu} \delta_{ij} \delta(x-y)
\eeq
and 
\beq 
\eta_{i,j}(x,y) = \frac{\mu^4}{{\rm g}} \delta_{ij} \delta(x-y)
\eeq
where ${\rm g}$ is the gauge coupling and $\mu$ is a constant of mass dimension one.
For Einstein gravity, the unregulated metrics are 
\bea
 \gamma^{\mu\nu, \rho\sigma }(x,y) &=&   \frac{\mu^{2}}{64 \pi G_N}  \sqrt{g} \left(g^{\mu\rho}g^{\nu \sigma} + g^{\mu\sigma}g^{\nu\rho} \right. \nn
 && -\left.   g^{\mu\nu} g^{\rho \sigma} \right)\delta(x-y)\,,
\eea
and
\begin{equation}\label{eq:eta}
\eta_{\mu, \nu}(x,y) = \frac{\mu^4}{16 \pi G} \sqrt{g} g_{\mu\nu} \delta(x-y)\,,
\end{equation}
respectively where $G_N$ is Newton's constant.  
Later, specifically in sections~\ref{sec:1YM} and ~\ref{sec:1Grav}, we will use these metrics and their inverse to raise and lower indices.

We consider  \eq{Z_can} the defining form of the functional integral since it refers in principle to a choice of section only and not to an auxiliary gauge fixing condition. In practice determining the explicit form of the integrand can only be achieved by a gauge fixing condition or giving an explicit expression for $\hat{\G}[\phi]$. However, one could proceed without choosing a specific section and just use the transformation property of the dressing field to compute observables without choosing a gauge. This approach is in line with seeing the dressing field as a technical choice. 
Alternatively, we could choose the dressing field in a physically motivated way which may suit a specific purpose by simplifying a calculation of desired observable.

\section{Relation to the Faddeev-Popov form}
\label{sec:FP}
Let us now derive the Faddeev-Popov (FP) form of the functional integral starting from our more fundemental form \eq{Z_can}. To do so we must assume the existence of some conditions  $\chi^{\a}[\phi] =0$  which do not have the Gribov ambiguity. This means that these conditions are equivalent to selecting a single representative for each orbit such that $\chi^{\a}[\g_*\phi] =0$, taken as an equation for $\g$, has a unique solution, namely
\beq
\chi^{\a}[\g_*\phi] =0 \implies \g = \hat{\G}[\phi]\,.
\eeq 
If this is the case we can relate the delta functions by
\beq
 \delta(\g- \hat{\G}[\phi]) = \delta(\chi[\g_*\phi]) \left| \det \frac{\delta \chi^\a[\g_*\phi]}{\delta \g^\b} \right|
\eeq
which when evaluated at $\g = \mathfrak{1}$ gives
\beq
 \delta(\mathfrak{1}- \hat{\G}[\phi]) = \delta(\chi[\phi]) \left| \det \frac{\delta \chi^\a[\g_*\phi]}{\delta \g^\b}  \right|_{\g = \mathfrak{1}}
\eeq
and thus demonstrates that from  \eq{Z_can} we can derive the FP form. Since we have shown already that the path integral is independent of the choice of representatives, we have shown independence of the choice of $\chi^{\a}[\phi]=0$ if there is no ambiguity.
However, note that if $\chi^{\a}[\phi]=0$ has an ambiguity we cannot go the other way, hence supporting our view that    \eq{Z_can} is more fundamental.

\section{Landau and harmonic gauges}
\label{sec:examples}
Since the notions of the dressed field and the corresponding dressing field play a central role. Here we will give two examples which correspond to the usual Landau and harmonic gauges in Yang-Mills and gravity respectivley.  
We can obtain dressed fields that obey these gauges by considering functionals $I[\phi, \hat{\g}]$ and minimise them with respect to the gauge transformations $\hat{\g}$ to obtain the corresponding dressing field $\hat{\G}$. For example in Yang-Mills one can minimise \cite{Zwanziger:1990tn}
\beq
I[{\rm a}, \hat{\g} ] = \int_x \tr (\hat{\g}_*{\bf a}_\mu) \,(\hat{\g}_*{\bf a}_\mu) \,,
\eeq
then the corresponding dressed gauge field is unique and obeys the Landau condition $\partial^{\mu} \hat{a}^i_\mu=0$.
For $SU(N)$ the group dressing field ${\bf D} = {\rm e}^{\im {\bf t}_i \hat{\G}^i}$ obeys
\beq \label{V_equation}
{\bf D}^\dagger  {\bf \nabla}_\mu {\bf \nabla}^\mu {\bf  D} + (\nabla_\mu {\bf D})^\dagger (\nabla_\mu{\bf D}) = 0
\eeq
where here $\nabla_\mu {\bf D} = \partial_\mu {\bf D} +  i  {\bf D} {\bf a}_\mu$ and the dagger denotes the comlpex conjugate. 
 
For gravity we can do something similar by minimising
\beq
I[g, \hat{\g} ]  = \int_x \g_* (\sqrt{g} g^{\mu\nu}) \delta_{\mu\nu}\, 
\eeq
then the corresponding dressed field obeys the harmonic gauge condition $  \partial_\mu (\sqrt{\hat{g}} \hat{g}^{\mu\nu})=0$.
Indeed, $I[g, \hat{\g} ]$ takes the form 
\beq
I[g, \hat{\g} ]  =  \int_x \sqrt{g}  g^{\rho \lambda} \delta_{\mu\nu}  \partial_\rho \hat{\g}^{\hat{\mu}}  \partial_{\lambda} \hat{\g}^{\hat{\nu}} \delta_{\hat{\mu}\hat{\nu}}   
\eeq
So the dressing fields which minimise $I[g, \hat{\g} ]$  obey the harmonic condition on the coordinates
\beq
\nabla^2_g \hat{\G}^{\hat{\mu}} = 0 
\eeq
where $\nabla^2_g$ is the Laplacian for a {\it scalar}.

With these examples in mind we can develop our approach keeping the choice of dressing field very general. 

\section{BRST action for  Yang-Mills}
\label{sec:YMaction}
In the case of Yang-Mills the path integral is given by 
\beq \label{Z_YM}
\mathcal{Z} =    \int d \phi \, \tilde{\mu}[\phi] \delta\left(\hat{\mathfrak{G}}[\phi]\right) e^{-S[\phi]}
\eeq
In order to write the delta function in the path integral in a convenient way for Yang-Mills we
are going to impose a constraint on the group dressing fields since the they transform linearly.
This leads to some complications that we do not encounter in gravity.

To this end we consider a smoothened delta function $\delta_{\eps}({\bf M} - \bf{1})$ on the space of complex matrices such that
 \beq
\lim_{\eps\to 0} \delta_{\eps}({\bf M} - \bf{1})   =  \delta({\bf M} - \bf{1})  
\eeq
 where as a distribution $\delta({\bf M} - \bf{1})$ only has support for ${\bf M} = \bf{1}$ hence 
\beq
 \delta({\bf M} - {\bf 1})  F({\bf M}) =  \delta({\bf M} - \bf{1})  F(\bf{1})\,.
\eeq
In particular let's take 
\beq
 \delta_{\eps}({\bf M} - {\bf 1}) = \int d B   e^{-\int_x \frac{1}{2} \left[  \eps \, \tr  {\bf \bar{B}} {\bf B} + \tr  {\bf \bar{B}} ( {\bf M} - \bf{1}) +    \tr ( {\bf \bar{M}} - \bf{1}) {\bf B}\right]}
\eeq
where ${\bf B}$ and  ${\bf \bar{B}}$ are the Nakanishi-Lautrup fields which we take to be complex matrices.
Here we strictly take ${\bf \bar{B}}$ to be the complex conjugate of ${\bf B}$ and thus the independent fields are the real and imaginary parts $B^{\Re}_{IJ} $ and $B^{\Im}_{IJ}$  respectively. We note that
\beq
B_{IJ} = B^{\Re}_{IJ} + \im B^{\Im}_{IJ}\,,
\eeq
and
\beq
\bar{B}_{IJ} = B^{\Re}_{JI} - \im B^{\Im}_{JI}\,,
\eeq
\\
which imply\,,
\beq
B^{\Re}_{IJ} = \frac{1}{2} \left( B_{IJ} + \bar{B}_{JI}\right)\,,
\eeq
and
\beq
B^{\Im}_{IJ} =- \frac{\im}{2} \left( B_{IJ} - \bar{B}_{JI}\right)\,.
\eeq
\\
Therefore we can work with ${\bf \bar{B}}$ and  ${\bf B}$ as if they are the independent fields for most calculations.

Now we consider the delta function on the group manifold $\delta(\g)$ which we can write as
\beq
\delta(\g) = \lim_{\eps\to 0}  N_{\eps} \,  \delta_{\eps}({\bf G}[\g] - {\bf1}) \,, 
\eeq
with 
\beq
N_{\eps} =    \frac{1}{\int d \h  \delta_{\eps}({\bf G}[\h] - \bf{1})}
\eeq
To see this we need to carefully think about our choice to parameterise the group by the $\g^i$ which are not strictly global coordinates on the gauge manifold (since none exist for e.g. $SU(2)$ theories the manifold is a 3-sphere). Instead for each element of the group there are multiple corresponding values of the parameters. 
Therefore to integrate over the group we can make a change of variables to some local coordinates $\psi^i$ on a coordinate patch that includes the identity and understand $\int d \g \dots $ as an integral over the coordinates $\psi^i$ on this patch. Then restricted to this patch
\beq
{\bf G}[\g] = {\bf1} \implies \g^i=0 
\eeq
from which is follows that 
\beq
\int d \g  \lim_{\eps\to 0}  N_{\eps} \, \mathcal{F}(\g)  \delta_{\eps}({\bf G}[\g] - {\bf1}) =  \mathcal{F}(0)  
\eeq

Recalling that ${\bf D}[\phi]$ are the group dressing fields, we can write 
\begin{widetext}
\beq
\delta(\hat{\G}[\phi]) =  \lim_{\eps\to 0} N_\eps   \int d B   e^{-\int_x \frac{1}{2} \left[  \eps  \, \tr  {\bf \bar{B}} {\bf B} + \tr  {\bf \bar{B}}( {\bf D}[\phi] - \bf{1}) +    \tr ( {\bf \bar{D}}[\phi] - \bf{1}) {\bf B}\right]}
\eeq
\end{widetext}
Furthermore in the limit $\eps \to 0$ the factor $N_{\eps}$ can be  written as an integral over Grassmann fields
\beq
N_\eps \sim  \int dC d A  e^{- \frac{1}{2} \int_x \tr \left[ \eps^2 {\bf \bar{A}}{\bf A} + \im {\bf \bar{A}} {\bf t}_i C^i - \im {\bf A} {\bf \bar{t}}_i C^i    \right] }
 \eeq
 This resembles the integration of FP ghosts although it gives just a constant factor.
 As we have said the FP determinant is one so we do not need to have ghosts in our path integral.
 Nonetheless to have BRST symmetry manifest  we can introduce the ghosts.
To this end let's define the FP operator
\beq
{\bf Q}\,_i(x,y) := - \left.  \frac{\delta}{\delta \g^i(y)} {\bf D}[\g_* \phi](x) \right|_{\g = \mathfrak{1}}\,.
\eeq
which gives
\beq
{\bf Q}\,_i(x,y)  = \im {\bf D} {\bf t}_i \delta(x-y)
\eeq
then we note that when the gauge condition, ${\bf D} \!!! {\bf 1}$, applies ${\bf Q}\,_i(x,y) \!!! \im {\bf t}_i \delta(x-y)$. 
Suppressing terms which vanish as $\eps \to 0$, this allows us to write the path integral as 
\beq
\mathcal{Z} =  \int d\phi dA dB   d C  \,  \tilde{\mu} \,     {\rm e}^{-\S +\frac{1}{2} \int_x (\tr {\bf B} + \tr  {\bf \bar{B}})}\,.
\eeq
where
\beq
\S = S+ \frac{1}{2}  \int_x \tr  \left[  {\bf \bar{B}} {\bf D}+      {\bf \bar{D}}{\bf B}  +( {\bf \bar{A}} {\bf Q}_i  + {\bf A} {\bf \bar{Q}}_i ) C^i    \right]\,.
\eeq
Here one should remember to give a (squared) mass $\eps$ to the fields to the complex matrix valued fields and take the limit where this mass goes to zero.  Note that the fields $C^i(x)$ are the ghosts, ${\bf A}(x)$ are the anti-ghosts (not to be confused with gauge fields which we denote as ${\bf a }_\mu$) and ${\bf B}(x)$ the Nakanishi-Lautrup fields.
Introducing a DeWitt notation
 \beq
D^{\hat{\a}} = ({\bf D})^{IJ}(x)
\eeq
we can write 
\beq
\S = S+ \int_x \frac{1}{2} \left[ B_{\ha}^*  D^{\ha}+    B_{\ha}  \bar{D}^{\ha}  + (A^*_\ha Q^\ha\,_\a  +  A_{\ha} \bar{Q}^{\ha}\,_{\a})
  C^\a    \right]\,.
\eeq
where the asterisk denotes the complex conjugation.

\section{BRST action for  gravity}
\label{Sec:BRSTactionGravity}
We now construct a BRST invariant action based on our gauge fixing conditions for gravity.
We adopt a similar notation to the Yang-Mills case by denoting the dressing fields as $D^{\hat{\a}}$ which in the case of gravity transform as a set of four scalar fields. The situation is actually simpler in gravity and we can write the path integral simply as
\beq 
\mathcal{Z} = \int d \phi  d B   \, \tilde{\mu}[\phi]\, e^{-S[\phi]- B_{\hat{\a}}  (D^{\hat{\a}}[\phi]-  \mathfrak{1}^{\hat{\a}})  }
\eeq
where we do not need to introduce any small mass $\eps$ in this case.
Then we define the FP operators as 
\beq
Q^{\hat{\mu}}\,_\mu(x,y) := - \left.  \frac{\delta}{\delta \g^\mu(y)} D^{\hat{\mu}}[\g_* \phi](x) \right|_{\g = \mathfrak{1}}\,.
\eeq
which gives
\beq
Q^{\hat{\mu}}\,_\mu(x,y) =   \partial_\mu D^{\hat{\mu}} \delta(x-y)
\eeq

When  $D^{\hat{\mu}}(x)= \mathfrak{1}^{\hat{\mu}} = x^{\hat{\mu}}$  we have that $Q^{\hat{\a}}\,_\b \!!! \delta^{\hat{\a}}_{\b}$. Therefore we can add a factor into the path integral a factor of the form
\beq
 \int dA  d C e^{-A_{\hat{\a}} Q^{\hat{\a}}\,_\b  C^\b}\,,
\eeq
where  $C$ and  $A$ are the ghost and anti-ghost. 
Writing
\beq \label{Total_action_1}
\S = S +   \int_x \left( A_{\hat{\mu}}(x) \partial_\mu D^{\hat{\mu}} C^{\mu}  +    B_{\hat{\mu}} D^{\hat{\mu}}\right) \,,
\eeq
or in DeWitt notation
\beq \label{Total_action}
\S = S +  A_{\hat{\a}} Q^{\hat{\a}}\,_\b C^\b +   B_{\hat{\a}}  D^{\hat{\a}} \,,
\eeq
where 
\beq
B_{\hat{a}} D^{\hat{\a}} =  \int_x B_{\hat{\mu}} D^{\hat{\mu}}\,,
\eeq
\beq
 A_{\hat{\a}} Q^{\hat{\a}}\,_\b C^\b = \int_x A_{\hat{\mu}}(x) \partial_\mu D^{\hat{\mu}} C^{\mu} \,,
\eeq
and 
\beq \label{B1gravity}
\mathfrak{1}^{\hat{\a}} B_{\hat{\a}} = \int_x x^{\hat{\mu}} B_{\hat{\mu}}(x)\,.
\eeq
The functional integral for gravity is therefore given by
\beq
\mathcal{Z} = \int d\phi   dA  d B   d C \,  \tilde{\mu}[\phi] e^{-\S[\phi,A,B,C] +\mathfrak{1}^{\hat{\a}} B_{\hat{\a}}}\,.
\eeq

\section{Symmetries}
\label{sec:Sym}
We now analyse the symmetries of the action $\S$ for Yang-Mills and for gravity.

\subsection{BRST invariance}
For both Yang-Mills and gravity the generators of gauge transformations $T^a_\a$ satisfy
\beq \label{Lie_Algebra}
[\vec{T}_{\alpha}, \vec{T}_{\beta}] = f^{\gamma}\,_{\alpha \beta} \vec{T}_{\gamma}\,,
\eeq
where $f^{\gamma}\,_{\alpha \beta}$ are the structure coefficients.
For Yang-Mills
\beq
 f^{\gamma}\,_{\alpha \beta} \xi^\alpha \eps^\beta =  f^{k}_{i j} \xi^i \eps^j\,.
 \eeq
For gravity
\beq
 f^{\gamma}\,_{\alpha \beta} \xi^\alpha \eps^\beta =   \mathcal{L}_{\xi} \eps^\mu =  \xi^\nu \partial_\nu \eps^\mu  -  \partial_\nu \xi^\mu \eps^\nu  \,.
\eeq
They obey
\beq
f^{\gamma}\,_{\a\b} = -f^{\gamma}\,_{\b\a} \,,
\eeq
and
\beq
 f^{\eps}\,_{\a\b} f^{\delta}\,_{\eps \gamma} +  f^{\eps}\,_{\b \c } f^{\delta}\,_{\eps \a} +  f^{\eps}\,_{\c \a } f^{\delta}\,_{\eps \b} = 0 \,.
\eeq
Furthermore we have that
\beq \label{QT}
Q^{\ha}\,_{\a} = - \frac{\delta  D^{\ha}}{\delta \phi^a} T^a_\a
\eeq  
Then it follows that the actions $\S$ are BRST invariant with 
\begin{equation}\label{eq:BRST}
\begin{array}{r@{}l} 
\delta_\theta \phi^a &{}= - \theta T^a_{\alpha}[\phi] C^{\alpha}  \,,  \vspace{5pt} \\
\delta_\theta C^{\alpha} &{}=  \theta \frac{1}{2}C^{\beta} f^{\alpha}_{\beta \gamma}  C^{\gamma}  \,,   \vspace{5pt} \\
\delta_\theta A_{\hat{\alpha}} &{}=   - \theta B_{\hat{\a}}  \,,
  \vspace{5pt} \\
\delta_\theta B_{\hat{\alpha}} &{}=  0 \,.
\end{array}
\end{equation} 
where $\theta$ is a  Grassmann parameter. 

One can check that the measure is also BRST invariant in particular the transformation properties of $\gamma_{ab}[\phi]$ and $\eta_{\a\b}[\phi]$ ensure this.

 For Yang-Mills we have 
\beq
\S = S + \frac{1}{2} \int_x \tr   \left[  {\bf \bar{B}} {\bf D}+      {\bf \bar{D}}{\bf B}  + \im ( {\bf \bar{A}} {\bf D} {\bf C}  - {\bf A} {\bf {\bf \bar{C} }\bar{D}} )     \right]
\eeq
with
\bea
{\bf C} \equiv C^i {\bf t}_i\\
{\bf \bar{C}} \equiv C^i {\bf \bar{t}}_i
\eea
Then it follows that $\S$ is BRST invariant with 
 \bea
\delta_\theta {\bf a}_\mu & =& \nabla_\mu^{\rm a} {\bf C} \theta \,, \\
\delta_\theta {\bf C} &=&  - \im {\bf C}^2\theta   \,, 
\eea
For gravity the BRST transformations of the metric and ghosts are given by
 \bea
\delta_\theta g_{\mu\nu} & =& -( \nabla_{\mu} C_{\nu}+  \nabla_{\nu} C_{\mu} )  \theta \,, \\
\delta_\theta C^{\mu} &=&  C^{\mu} \nabla^{g}_{\nu} C^{\nu} \theta = C^{\mu} \partial_{\nu} C^{\nu} \theta   \,. 
\eea
\subsection{Gauge invariance}
Remarkably, as well as being BRST invariant, $\S$ is gauge invariant in the case of Yang-Mills.
With the gauge transformations extend to the other fields: 
\bea
\g_*{\bf B} &=& {\bf  B}{\bf \bar{G}}[\g]\\
\g_*{\bf A} &=& {\bf  A}{\bf \bar{G}}[\g]\\
\g_*{\bf C} &=&  {\bf  G}[\g]  {\bf  C}{\bf G}^{-1}[\g]
\eea
which imply
\bea
\g_*{\bf \bar{B}} &=&{\bf G}[\g] {\bf  B}\\
\g_*{\bf \bar{A}} &=& {\bf G}[\g] {\bf  A}\\
\g_*{\bf \bar{C}} &=&  {\bf  \bar{G}}^{-1}[\g]  {\bf  \bar{C}}{\bf \bar{G}}[\g]\\
\g_* C^i &=& ({\rm e}^{F(\g)})^i\,_j C^j
\eea
Note however that $\frac{1}{2} \int_x (\tr {\bf B} + \tr  {\bf \bar{B}})$ is not gauge invariant but is BRST invariant.

\subsection{Diffeomorphism invariance}
 For gravity the action $\S$ is diffeomorphism invariant  where  we define
\bea
\g_* C^\mu(x) &=&  \frac{\partial \g^\mu(x)}{\partial x^\nu}  C^{\nu}(\hat{\g}(x))\\
\g_* A_{\hat\mu}(x) &=&  \det \left|  \frac{\partial \hat{\g}(x)}{\partial x}   \right|   A_{\hat{\mu}}(\hat{\g}(x))\\
\g_* B_{\hat\mu}(x) &=&  \det \left|  \frac{\partial \hat{\g}(x)}{\partial x}   \right|   B_{\hat{\mu}}(\hat{\g}(x))
\eea
from which we observe that $C^\mu(x)$ transforms as a vector and both  $A_{\hat\mu}(x)$ and $B_{\hat\mu}(x)$ transform as four scalar densities of weight one.

\subsection{Gauge Invariance of the BRST measure}
One can also show that the BRST measure involving the ghosts, anti-ghosts and Nakanishi-Lautrup fields,  is gauge invariant in Yang-Mills and diffeomorphism invariant in gravity under {\it field independent} gauge transformations.
  Indeed separately we have 
\beq
 d(\g_*\phi) \sqrt{\det \gamma[\g_*\phi]} =  d\phi \sqrt{\det \gamma[\phi]} \,,
\eeq
\beq
d (\g_*C ) \frac{1}{\sqrt{\det \eta[\g_* \phi] } }=  d C  \frac{1}{\sqrt{\det \eta[\phi]} }\,,
\eeq
and 
\beq
 dA  dB =  d(\g_*A)  d(\g_*B)\,.
\eeq
It is important that this invariance is both linear with the transformations themselves are independent of the fields (unlike exchangers which must depend on the fields).  Gauge invariance and diffeomorphism invariance are therefore broken in the functional integral only by the factors ${\rm e}^{\int_x x^{\hat{\mu}} B_{\hat{\mu}}}$ and  ${\rm e}^{\frac{1}{2} \int_x (\tr {\bf B} + \tr  {\bf \bar{B}})}$  respectively.
 
\section{Effective action}
\label{sec:EA}
Let's now introduce sources into the functional integral for each field. For the Nakanishi-Lautrup fields we will do this by replacing the terms which break gauge invariance $S_{\rm id}[B]$ by a source term. For Yang-Mills we replace
\bea \label{SidYM}
S_{\rm id}[B]\equiv \frac{1}{2} \int_x \tr ( {\bf B} +   {\bf \bar{B}}) &\to& \int_x  \tr ( {\bf \bar{K}} {\bf B} + \tr{\bf K} {\bf \bar{B}})\\
&=&2 \int_x  \left( B^{\Re}_{IJ} K^{IJ}_{\Re}+ B^{\Im}_{IJ} K^{IJ}_{\Im} \right) \nonumber
\eea
where ${\bf K} = {\bf K}^{\Re} + \im  {\bf K}^{\Im}$ is a complex source. 
Similarly for gravity we replace the terms that break diffeomorphism invariance according to 
\beq \label{SidGR}
S_{\rm id}[B]\equiv \int_x x^{\hat{\mu}} B_{\hat{\mu}} \to \int_x K^{\hat{\mu}} B_{\hat{\mu}} 
\eeq
In this way we absorb the terms that break the symmetries into the source terms.

To write things compactly let us introduce a super-field notation for the fields and the sources such that $\Phi^A$ collects all the fields components, i.e. $\{ \Phi\} =\{ \phi, A, B,C \}$,  and  $\mathcal{J}_A$ collects the sources for each field i.e. for the original fields, the ghosts, the anti-ghosts and  Nakanishi-Lautrup fields.
Then we define the Schwinger function by
\bea \label{Schwinger}
&&e^{W[\mathcal{J}]} = \int d\Phi \, \tilde{\mu}[\Phi] e^{-\S[\Phi] +  \mathcal{J}_A  \Phi^A }  
\eea
and the effective action as its Legendre transform 
\beq \label{Gamma}
\Gamma[\Phi] = \sup\nolimits_\mathcal{J}  \mathcal{J}_A \Phi^A - W[\mathcal{J}] \,,
\eeq
where $\Phi^A$ is now denoting the expectation value of the fields in this context\footnote{We will use $\Phi$ both for the mean field and the integration variable when it is clear from the context which field we are referring to. When can confusion can arise, i.e. when the we refer to the quantum field as an operator, we but a prime on the integration variable such that $\langle \Phi'\rangle = \Phi$.} i.e. 
\beq \label{Phi_to_J}
\Phi^A = \frac{\delta W[\mathcal{J}]}{\delta \mathcal{J}_A} \equiv U^{A}[\mathcal{J}]\,.
\eeq
We can solve this equation to obtain $\mathcal{J}_A =  V_A[\Phi]$.
Then the meaning of \eq{Gamma} is that the rhs is evaluated at its supremum, namely we have that
\beq \label{GammaPhi}
\Gamma[\Phi]  = V_A[\Phi] \Phi^A - W[V[\Phi]]
\eeq
It follows that
\beq \label{Gamma1}
\mathcal{J}_A =  V_A[\Phi] =   \Gamma[\Phi]   \frac{\overleftarrow{\delta}   }{\delta \Phi^A}   \,
\eeq
where the notation means the derivative acts to the left.
Furthermore we have that that $V[\cdot]$ and $U[\cdot]$ are inverse maps:
\beq
\mathcal{J}_A  = V_A[U[\mathcal{J}]] = \mathcal{J}_A\,,   \,\,\,\,\,\,  U^A[V[\Phi]] = \Phi^A
\eeq

One can then compute correlation functions from $\Gamma[\Phi]$ in the standard way. For example denoting $P^{AB}[\Phi]$ as the inverse of the Hessian of $\Gamma[\Phi]$ we have that 
\beq
P^{AB}[\Phi] = \langle \Phi'^A  \Phi'^B \rangle_{\mathcal{J}} - \langle \Phi'^A \rangle  \langle \Phi'^B \rangle_{\mathcal{J}}
\eeq
where the subscript indicates the expectation value is in the presence of the source terms.
\\

Let's define $\bar{\Phi}$ as a solution to 
\bea
\frac{\delta \Gamma}{\delta \phi^a}  = 0\,, \,\,\,\,\,  \frac{\delta \Gamma}{\delta A_{\hat{\a}}} = 0  \,  \,\,\,\,\,
\frac{\delta \Gamma}{\delta C^\a}   = 0\,,    
\eea
along with 
\beq \label{Gamma1YM}
\frac{\delta \Gamma}{\delta B_{IJ}^{\Re}(x)} = \delta_{IJ}\,, \,\,\,\,\,\,  \frac{\delta \Gamma}{\delta B_{IJ}^{\Im}(x)} = 0
\eeq
for Yang-Mills, and
\beq \label{Gamma1Gravity}
 \frac{\delta \Gamma}{\delta B_{\hat{\mu}}(x)} = x^{\hat{\mu}}
\eeq
for gravity. Which we summarise for both theories as  
\beq \label{Gamma1}
 \frac{\delta \Gamma}{\delta B_{\hat{\a}}(x)} = {\bf 1}^{\hat{\a}}\,.
\eeq

Consequently, by \eq{Gamma1} all sources are zero apart from the ones needed to implement the delta function in the path integral by imposing \eq{Gamma1YM} and \eq{Gamma1Gravity}. Since the delta functions impose that $\phi' = \hat{\phi}[\phi']$ all correlation functions are correlation functions of the dressed fields. For example we have that
\beq
\bar{\phi}^a = \langle \hat{\phi}^a[\phi'] \rangle \,,
\eeq
is the expectation value of the dressed field
and that 
\beq
P^{ab}[\bar{\Phi}] =  \langle \hat{\phi}^a[\phi']  \hat{\phi}^b[\phi'] \rangle - \langle \hat{\phi}^a[\phi'] \rangle  \langle \hat{\phi}^b[\phi'] \rangle\,,
\eeq
is the connected two point function of dressed fields.

\section{Gauge and diffeomorphism invariance of the effective action}
\label{sec:Inv}
We now show that our effective actions are gauge (and diffeomorphism) invariant.
First let's summarise the gauge transformations in super-field notation by 
\beq
\g_*\Phi^A = T^A\,_B[\g] \Phi^B  + T^A[\g]
\eeq
so $T^A\,_B[\g]$ collects all the components of the gauge transformations and  $T^A[\g]$ is non-zero for the gauge field in Yang-Mills.
We then define the action of the gauge transformations on the source by
\beq
\g_* \mathcal{J}_A =  \mathcal{J}_B T^B\,_A[\hat{\g}] \,.
\eeq
By making a change of integration variables $\Phi \to \hat{\g}_* \Phi$ in the integral \eq{Schwinger} and using the invariance of the action $\S$ and the measure one has that
 \beq
 W[\g_* \mathcal{J}] = W[\mathcal{J}] - T^A[\hat{\g}] \mathcal{J}_A
 \eeq
 where we have used \eq{Tinverse1} and \eq{Tinverse2}.
 Differentiating this with respect to $\mathcal{J}$ on both sides  and using \eq{Phi_to_J} gives
 \beq
 \g_*U^A[\mathcal{J}] = U^{A}[\g_*\mathcal{J}]
 \eeq
 where $\g_*$ acts on $U^A$ as it does on $\Phi^A$.
 So we have
 \beq
V_{A}[\g_*U[\mathcal{J}]]   =  V_{A}[U[\g_*\mathcal{J}]]= \g_*\mathcal{J}
 \eeq
 and setting $\mathcal{J} \to V_A[\Phi]$ we get 
 \beq
 V_{A}[\g_*\Phi] = \g_* V_A[\Phi]
 \eeq
 where $\g_*$ acts on $V_A$ as it does on $J_A$.
 Then, using the expression  \eq{GammaPhi} and the previously obtained identities above, it follows that $\Gamma[\Phi]$ is gauge invariant
\beq
\Gamma[\g_*\Phi]= \Gamma[\Phi]  \,.
\eeq
 Let us stress that this property is a {\it property of the effective action} in a particular class of gauges and should not be confused with {\it gauge invariance of the theory} which is instead the statement that observables are independent of how we fix the gauge.  
 
 Let us now explore some consequences of this symmetry.
 First, it follows that the action, obtained from $\Gamma[\Phi]= \Gamma[\phi, A,B,C]$ by putting the unphysical fields to zero, namely
 \beq
 \Gamma[\phi]\equiv\Gamma[\phi, 0,0,0]\,,
 \eeq
 is itself gauge invariant 
 \beq
\Gamma[\g_*\phi]= \Gamma[\phi]  \,.
\eeq
 By the conversation of ghost number
 \beq
\left( C^{\a} \frac{\delta}{\delta C^{\a}}  -  A_{\hat{\a}} \frac{\delta}{\delta A_{\hat{\a}}}  \right) \Gamma[\phi,A,B,C] =0 
 \eeq
 The equations of motion for the ghosts and anti-ghosts will be satisfied if both of these fields vanish. 
 Then we can define the effective dressing field by 
 \beq
\mathcal{D}^{\hat{\a}}[\phi] \equiv \frac{\delta \Gamma}{ \delta B_{\hat{\a}}}[\phi,0,0,0]  \,. 
 \eeq
 Then if we can solve (c.f. \eq{Gamma1})
 \beq \label{Effective_gauge}
 \mathcal{D}^{\hat{\a}}[\phi] = {\bf 1}^{\hat{\a}}
 \eeq
  along with the equations of motion for $\Gamma[\phi]$ then we go on-shell. 
  Put differently \eq{Effective_gauge}, should be a gauge condition for the mean fields picking out a unique solution to the effective equations of motion
  \beq
  \frac{\delta \Gamma[\phi]}{\delta \phi} = 0 \,.
  \eeq

\section{Tree-level and one-loop analysis of $SU(N)$ Yang-Mills}
\label{sec:1YM}
In this section we will compute the on-shell tree-level propagators and one-loop effective action. 
For simplicity we will consider  $SU(N)$ Yang-Mills theories. Then 
\beq
\bar{D}^{\hat{\a}} = {\bf \bar{D}}^{IJ} =  ({\bf D}^{-1})^{IJ}
\eeq
Then let's introduce 
\beq
Y_i^{IJ}(x,y) = \im t_i^{IJ}\delta(x-y)
\eeq
\beq
Y_{IJ}^i(x,y) = - 2 \im t^{JI}_i \delta(x-y)
\eeq
Then in condensed notation we have that
\beq
Y_{\a}^{\hat{\a}} Y_{\hat{\a}}^\b=  \delta_\a^\b\,,   \,\,\,\,\,\,\,\,\,\,  Y_{\hat{\a}}^\a  Y_{\a}^{\hat{\b}} = \Xi^{\hat{\b}}_{\perp}\,_{\hat{\a}}
\eeq
where
\beq
\Xi^{IJ}_{\perp}\,_{KL}(x,y) = 2  t^{IJ}_i  t^{i}_{LK} \delta(x-y)
\eeq
is a projector on to trace-free matrices.

We also introduce 
\beq
\sigma^{\ha \hb} = \sigma^{IJ,KL}(x,y) = - \frac{{\rm g} }{2\mu^4} \delta^{IL} \delta^{JK} \delta(x-y)
\eeq
which maps a field matrix to its transpose times a negative factor  i.e.
\beq
\sigma^{\ha \hb} M_{\hb} = -  \frac{{\rm g} }{2\mu^4}  (M^{\rm T})^{\ha}
\eeq
we denote its inverse by $\sigma_{\ha \hb}$ (which proportional to $\sigma^{\ha \hb}$). 
The factor $-{\rm g}/(2\mu^4)$ makes ensures that 
\beq
\sigma_{\hb \ha}Y_{\a}^{\hat{\a}}  =Y_{\a\hat{\b}} = \eta_{\a\b} Y^\b_{\hat{\b}} \,.
\eeq
It is then convenient to use $\sigma_{\ha \hb}$ to raise and lower greek hatted indices then $\Xi^{\ha\hb}_\perp = \Xi^{\ha}_\perp\,_\hc  \sigma^{\hc \hb}$ and   $ \Xi_{\ha\hb}^\perp = \sigma_{\ha \hc} \Xi^{\hc}_\perp\,_{\hb} $  are both symmetric.  Let's define
\beq
T^a_{\ha} := Y^{\a}_{\ha} T^a_\a 
\eeq
and
\beq
D_a^{\hat{\a}} = \frac{\delta D^{\hat{\a}}}{\delta \phi^a}\,, \,\,\,\,\,\,\,\,\,\,\,  \bar{D}_a^{\hat{\a}} = \frac{\delta \bar{D}^{\hat{\a}}}{\delta \phi^a}
\eeq
Then when ${\bf D}[\phi] = {\bf 1}$ we have that
\beq
 \bar{D}_a^{\hat{\a}}  = -  D_a^{\hat{\a}} 
\eeq
and that
\beq
D_a^{\hat{\a}} T^{a}_\a = - Y^{\hat{\a}}_\a \,, 
\eeq
or equivalently 
\beq
D_a^{\hat{\a}} T^{a}_\hb = - \Xi^{\hat{\a}}_\perp\,_\hb \;. 
\eeq

\subsection{Tree-level Propagators}
We consider the propagator putting the ghosts and Nakanishi-Lautrup fields to zero and assume we satisfy the equations of motion 
\beq
\frac{\delta S}{\delta \phi^a} = 0
\eeq
 and the gauge condition
\beq
{\bf D}[\phi] = {\bf 1}\,.
\eeq
 Then the  $\phi$-${\bf B}$-${\bf \bar{B}}$ components of the Hessian of $\S$ are
\beq
\S^{(2)}_{\rm phy} = \begin{pmatrix}
S_{ab}  \,\,\,\,\, &  \frac{1}{2} D^{\hat{\b}}_{a}  \,\,\,\,\, & -\frac{1}{2}  D^{\hat{\b}}_{a} \\ \\
 \frac{1}{2} D^{\hat{\a}}_{b}  \,\,\,\,\, & 0   \,\,\,\,\, & - \eps \sigma^{\ha \hb} \\\\
 -\frac{1}{2} D^{\hat{\a}}_{b}  \,\,\,\,\, & - \eps \sigma^{\ha \hb}  \,\,\,\,\, & 0 
\end{pmatrix}
\eeq
where
\beq
 S_{ab} \equiv \frac{\delta^2 S}{\delta \phi^a \delta \phi^b}\,,
 \eeq
and we must take the limit $\eps \to 0$. We can write the propagator as 
\beq
P_{\rm phy}  = \begin{pmatrix}
P^{bc} \, &  P_{\hat{\c}}^{b}  \, &   \bar{P}_{\hat{\c}}^{b} \\ \\
 P_{\hat{\b}}^{c}\, & P^{\ha\hb} \,& R^{\ha\hb} \\\\
 \bar{P}_{\hat{\b}}^{c}\, & R^{\ha\hb} \, & \bar{P}^{\ha\hb} 
\end{pmatrix}\;.
\eeq
Hence
\beq
 \begin{pmatrix}
S_{ab} \,\,\, &  \frac{1}{2} D^{\hat{\b}}_{a}  \,\,\, & - \frac{1}{2}  D^{\hat{\b}}_{a} \\ \\
 \frac{1}{2} D^{\hat{\a}}_{b}\,\,\, & 0 \,\,\,& - \eps \sigma^{\ha \hb} \\\\
 -\frac{1}{2} D^{\hat{\a}}_{b}\,\,\, & - \eps \sigma^{\ha \hb} \, & 0 
\end{pmatrix} 
\begin{pmatrix}
P^{bc} \, &  P_{\hat{\c}}^{b}  \, &   \bar{P}_{\hat{\c}}^{b} \\ \\
 P_{\hat{\b}}^{c}\, & P_{\hb\hc} \,& R_{\hb\hc} \\\\
 \bar{P}_{\hat{\b}}^{c}\, & R_{\hb\hc} \, & \bar{P}_{\hb\hc} 
\end{pmatrix} = 
\begin{pmatrix}
\delta^a_c \, & 0  \, & 0 \\ \\
0 \, & \delta^{\ha}_\hc \,& 0 \\\\
0\, & 0 \, &  \delta^{\ha}_\hc\;.
\end{pmatrix}
\eeq
We note that from the gauge invariance of $S$, when the equations of motion are satisfied, we have
 \beq
 S_{ab} T^b_\a = 0
 \eeq
Then, one then finds after a little work that 
\beq
P^{a}_\ha = - \bar{P}^a_\ha = - T^a_{\ha}
\eeq
\beq
P_{\ha\hb}=  \bar{P}_{\ha\hb} = - \frac{1}{2\eps} \Xi^{\perp}_{\ha\hb}
\eeq
\beq
R_{\ha\hb} = - \frac{1}{\eps} \sigma_{\ha \hb} + \frac{1}{2\eps} \Xi^{\perp}_{\ha\hb}
\eeq
Additionally $P^{ab}$ obeys
\beq
S_{ac} P^{cb} = \Pi^b_\perp\,_a \equiv \delta^a_b + T^a_\ha D^{\ha}_b\,,
\eeq
and 
\beq
D^{\ha}_b P^{bc} = \eps P^{c \ha} =  - \eps  T^{c \ha}\,.
\eeq
Writing 
\beq
\Box_{ab} = S_{ab} +  D^{\ha}_a D_{\ha b}\,,
\eeq
and $p^{ab}$ as the propagator (i.e. inverse) of $\Box_{ab}$.
We then have that 
\beq
\Box_{ab} T^b_\ha = -  D_{a\ha} \,,
\eeq
and 
\beq
p^{ab} D_b^\ha = -   T^{a\ha}\,, 
\eeq
\\
Let's define $P^{ab}_{\perp} =   P^{ab}|_{\eps =0}$ then  
\beq
 P^{ab}_\perp =  p^{ab} -  T_{\ha}^a \sigma^{\ha \hb} T^{b}_\hb = p^{ab} - T^a_\a T^{\a}_b\,.
 \eeq
 \\
For the ghosts and anti-ghosts we have $C$-${\bf A}$-${\bf \bar{A}}$ components of the Hessian
\beq
\S^{(2)}_{\rm gh} =
\begin{pmatrix}
0 \,\,\, &  \frac{1}{2} Y_\a^{\ha}  \,\,\, & - \frac{1}{2}  Y_\a^{\ha} \\ \\
 \frac{1}{2}  Y_\a^{\ha}  \,\,\, & 0 \,\,\,& - \eps \sigma^{\ha \hb}  \\\\
 - \frac{1}{2} Y_\a^{\ha} \,\,\, & - \eps \sigma^{\ha \hb}  \,\,\, & 0
\end{pmatrix} \,,
\eeq
which has the propagator
\beq
P_{\rm gh} =
\begin{pmatrix}
0 \, &  Y^\a_{\ha}  \, & -   Y^\a_{\ha} \\ \\
 Y^\a_{\ha}  \, & P_{\ha \hb} \,& R_{\ha \hb}  \\\\
 -  Y^\a_{\ha} \, & R_{\ha \hb}  \, & P_{\ha \hb}
\end{pmatrix}\,.
\eeq

\subsection{One-loop}

\beq
e^{-\Gamma} = e^{-S}   \frac{\sqrt{\det \gamma_{ab}} }{ \sqrt{\det \eta_{\a\b}}}   \frac{ \sqrt{\det \S^{(2)}_{\rm gh}} }{\sqrt{  \det \S^{(2)}_{\rm phy} } }
\eeq
 
\begin{widetext}
Taking into account the measure (i.e. $\gamma$ and $\eta$) merely changes the positions of some indices.
We then have that
\bea
\det \begin{pmatrix}
S^a\,_b \,\,\, &  \frac{1}{2} D^{\hat{\b} a}  \,\,\, & - \frac{1}{2}  D^{\hat{\b} a} \\ \\
 \frac{1}{2} D_{\hat{\a}b}\, & 0 \,&  -\eps \delta_{\ha}^\hb \\\\
 -\frac{1}{2} D_{\hat{\a}b}\, & - \eps \delta^{\ha}_\hb \, & 0 
\end{pmatrix}  &=& 
\det \begin{pmatrix}
p^a\,_b \, & 0\, & 0 \\\\
0\, & \delta^\ha_\hb\, &0\\\\
0 \, & 0\,& \delta^\ha_\hb\,
\end{pmatrix}  
\det \begin{pmatrix}
S^a\,_b \, &  \frac{1}{2} D^{\hat{\b} a}  \,\,\, & - \frac{1}{2}  D^{\hat{\b} a} \\ \\
 \frac{1}{2} D_{\hat{\a}b}\, & 0 \,&  -\eps \delta^{\ha}_\hb \\\\
 -\frac{1}{2} D_{\hat{\a}b}\, & - \eps \delta^{\ha}_\hb \, & 0 
\end{pmatrix} 
\det \Box^a\,_b\nn
&=& 
\det\begin{pmatrix}
\Pi^a_\perp\,_b \,\,\, &  - \frac{1}{2} T^{\hat{\b} a}  \,\,\, &  \frac{1}{2}  T^{\hat{\b} a} \\ \\
 \frac{1}{2} D_{\hat{\a}b}\,\,\, & 0 \,\,\,& - \eps \delta^{\ha}_\hb \\\\
 -\frac{1}{2}  D_{\hat{\a}b}\,\,\, &  -\eps \delta^{\ha}_\hb \,\,\, & 0 
\end{pmatrix} 
\det \Box^a\,_b\nn
\eea

One can then show that in the limit $\eps \to 0$ 
\beq
 \det\begin{pmatrix}
\Pi^a_\perp\,_b \,\,\, &  - \frac{1}{2} T^{\hat{\b} a}  \,\,\, &  \frac{1}{2}  T^{\hat{\b} a} \\ \\
 \frac{1}{2} D_{\hat{\a}b}\,\,\, & 0 \,\,\,& - \eps \delta^{\ha}_\hb \\\\
 -\frac{1}{2}  D_{\hat{\a}b}\,\,\, & - \eps \delta^{\ha}_\hb \,\,\, & 0 
\end{pmatrix}  \sim
 \det
\begin{pmatrix}
0 \,\,\, &  \frac{1}{2} Y_\a^{\ha}  \,\,\, & - \frac{1}{2}  Y_\a^{\ha} \\ \\
 \frac{1}{2}  Y_{\a \ha}  \,\,\, & 0 \,& - \eps \delta^{\ha \hb}  \\\\
 - \frac{1}{2} Y_{\a \ha} \,\,\, & - \eps \delta^{\ha \hb}  \,\,\, & 0
\end{pmatrix}
\eeq
such that the FP determinant is cancelling this determinant leaving just $\det \Box^a\,_b$.

To see this we can check that with $\eps =0$ that the product of non-zero eigenvalues are equal.
 Explicitly we have 
\beq
\begin{pmatrix}
\Pi^a_\perp\,_b \,\,\, &  - \frac{1}{2} T^{\hat{\b} a}  \,\,\, &  \frac{1}{2}  T^{\hat{\b} a} \\ \\
 \frac{1}{2} D_{\hat{\a}b}\,\,\, & 0 \,\,\,&0 \\\\
 -\frac{1}{2}  D_{\hat{\a}b}\,\,\, &0  \,\,\, & 0 
\end{pmatrix} 
\begin{pmatrix}
\sqrt{2} T^b_\a v^\a \,\,\,  \\ \\
-Y_{\hb \a} v^{\a} \\\\
 Y_{\hb \a }  v^{\a}
\end{pmatrix} =  \frac{1}{\sqrt{2}} \begin{pmatrix}
\sqrt{2} T^a_\a v^\a \,\,\,  \\ \\
- Y_{\ha \a} v^{\a} \\\\
   Y_{\ha \a }  v^{\a}
\end{pmatrix}
\eeq
\beq
\begin{pmatrix}
\Pi^a_\perp\,_b \,\,\, &  - \frac{1}{2} T^{\hat{\b} a}  \,\,\, &  \frac{1}{2}  T^{\hat{\b} a} \\ \\
 \frac{1}{2} D_{\hat{\a}b}\,\,\, & 0 \,\,\,&0 \\\\
 -\frac{1}{2}  D_{\hat{\a}b}\,\,\, &0  \,\,\, & 0 
\end{pmatrix} 
\begin{pmatrix}
-\sqrt{2} T^b_\a v^\a \,\,\,  \\ \\
-Y_{\hb \a} v^{\a} \\\\
 Y_{\hb \a }  v^{\a}
\end{pmatrix} = - \frac{1}{\sqrt{2}} \begin{pmatrix}
-\sqrt{2} T^a_\a v^\a \,\,\,  \\ \\
- Y_{\ha \a} v^{\a} \\\\
   Y_{\ha \a }  v^{\a}
\end{pmatrix}
\eeq
and with $D_{a}^\a v^a_{\perp} = 0$
\beq
\begin{pmatrix}
\Pi^a_\perp\,_b \,\,\, &  - \frac{1}{2} T^{\hat{\b} a}  \,\,\, &  \frac{1}{2}  T^{\hat{\b} a} \\ \\
 \frac{1}{2} D_{\hat{\a}b}\,\,\, & 0 \,\,\,&0 \\\\
 -\frac{1}{2}  D_{\hat{\a}b}\,\,\, &0  \,\,\, & 0 
\end{pmatrix} 
 \begin{pmatrix}
v^b_{\perp}  \\ \\
0 \\\\
0
\end{pmatrix}
= 
 \begin{pmatrix}
v^a_{\perp}  \\ \\
0 \\\\
0
\end{pmatrix}
\eeq
so these eigenvalues are all equal to one and do not contribute to the product.
For the ghosts determinant we have the eigenvectors
\beq
\begin{pmatrix}
0 \,\,\, &  \frac{1}{2} Y_\a^{\ha}  \,\,\, & - \frac{1}{2}  Y_\a^{\ha} \\ \\
 \frac{1}{2}  Y_{\a \ha}  \,\,\, & 0 \,&  0  \\\\
 - \frac{1}{2} Y_{\a \ha} \,\,\, & 0  \,\,\, & 0
\end{pmatrix}
\begin{pmatrix}
 \sqrt{2} v^\b \,\,\,  \\ \\
 Y_{\hb \a} v^{\a} \\\\
-   Y_{\hb \a }  v^{\a}
\end{pmatrix}
= \frac{1}{ \sqrt{2} } \begin{pmatrix}
 \sqrt{2}  v^\a \,\,\,  \\ \\
 Y_{\ha \a} v^{\a} \\\\
-   Y_{\ha \a }  v^{\a}
\end{pmatrix}
\eeq
and
\beq
\begin{pmatrix}
0 \,\,\, &  \frac{1}{2} Y_\a^{\ha}  \,\,\, & - \frac{1}{2}  Y_\a^{\ha} \\ \\
 \frac{1}{2}  Y_{\a \ha}  \,\,\, & 0 \,&  0  \\\\
 - \frac{1}{2} Y_{\a \ha} \,\,\, & 0  \,\,\, & 0
\end{pmatrix}
\begin{pmatrix}
- \sqrt{2} v^\b \,\,\,  \\ \\
 Y_{\hb \a} v^{\a} \\\\
-   Y_{\hb \a }  v^{\a}
\end{pmatrix}
= - \frac{1}{ \sqrt{2} } \begin{pmatrix}
- \sqrt{2}  v^\a \,\,\,  \\ \\
 Y_{\ha \a} v^{\a} \\\\
-   Y_{\ha \a }  v^{\a}
\end{pmatrix}
\eeq
\end{widetext}
The other eigenvalues that vanish for $\eps \to 0$ are just equal to $-\eps$ with eigenvectors $\{ 0 ,v^{\ha}, v^{\ha}\}^T$ in both cases.

So we arrive at 
\beq
e^{-\Gamma} = e^{-S} \frac{ 1  }{ \sqrt{ \det \Box^a\,_b  } } =  e^{-S}  \sqrt{ \det p^a\,_b  } 
\eeq
Using the identity for $p^a\,_b$ we find that 
\beq
e^{-\Gamma} = e^{-S} \sqrt{ \frac{\det T^{\a}_{b} T^b_\b   }{  \det\nolimits_\perp S^a\,_b   } }
\eeq
where $\det\nolimits_\perp$ specifies that we take the non-zero eigenvalues and we have used that the longitudinal eigenvalues of $p^a\,_b$ are same as those of $T^{\a}_{b} T^b_\b$. This expression is independent of the gauge and agrees with the standard result. Indeed an equivalent form is 
\beq
e^{-\Gamma} = e^{-S} \frac{\det T^{\b}_{b} T^b_\c   }{   \sqrt{ \det S^a\,_b + T^{a}_\a T^{\a}_b   } }
\eeq
which is the form in a minimal background field gauge i.e. $\chi^\a =  T_a^\a[\phi_0] \phi^a$ where $\phi_0$ is the background field. We stress however that we have reached this result without any background field and that in our gauge the ghosts only contributed a constant factor.
\\

\section{Tree-level and one-loop analysis of gravity}
\label{sec:1Grav}
Now we repeat the analysis for gravity.
Again we put the ghosts and Nakanishi-Lautrup fields to zero and assume we satisfy the equations of motion 
\beq
\frac{\delta S}{\delta \phi^a} = 0\,,
\eeq
 and the gauge condition
 \beq
  D^{\hat{\mu}}(x) =  x^{\hat{\mu}} \,.
 \eeq

\subsection{Tree-level propagator}
If $P^{ab}$, $P^{a\ha}$ and $P^{\ha\hb}$ are the components of the propagator in the $\phi$-$B$ sector then we must have that at tree-level
\begin{widetext}
\bea
\label{PropEquation}
\begin{pmatrix}
S_{ab} \, & D^{\hat{\b}}_{a} \\
 D^{\hat{\a}}_{b}\, & 0 
\end{pmatrix}
\begin{pmatrix}
P^{bc} \, & P_{\hat{\c}}^{b} \\
 P_{\hat{\b}}^{c}\, & P_{\hat{\b}\hat{\c}} 
\end{pmatrix}
=
\begin{pmatrix}
S_{ab}P^{bc} +   D^{\hat{\b}}_{a}   P_{\hat{\b}}^{c} \,\,\, & S_{ab} P_{\hat{\c}}^{b} +  D^{\hat{\b}}_{a}  P_{\hat{\b}\hat{\c}}\\
 D^{\hat{\a}}_{b} P^{bc}      \,\,\, &  D^{\hat{\a}}_{b}  P_{\hat{\c}}^{b}
\end{pmatrix}
= \begin{pmatrix}
\delta^a_c\, & 0 \\
0 \, & \delta^{\hat{\a}}\,_{\hat{\c}} 
\end{pmatrix}
\eea
\end{widetext}

So we have four equations that must be satisfied by the propagator, corresponding to the four components of the matrix equation \eq{PropEquation}.  We solve the bottom right corner with 
\beq
P_{\hat{\a}}^a = - T^a_{\hat{\a}} 
\eeq
which follows from \eq{QT} with $Q^{\hat{\a}}\,_\b = \delta^{\hat{\a}}_\b$.
Then by gauge invariance $S_{ab} T^b_\a=0$ and so  $S_{ab} P_{\hat{\c}}^{b} = 0$. Therefore we can set $P_{\hat{\b}\hat{\c}} = 0$ to solve the top right corner. Furthermore,
\beq
  P_{\hat{\b}}^{a}  D^{\hat{\b}}_{b}  = \Pi_{\parallel}^a\,_b \equiv - T^a_{\hat{\a}} D^{\hat{\a}}_b  
\eeq
is the longitudinal projector and we have from the top left corner
\beq \label{S2P}
S_{ab}P^{bc} = \Pi_{\perp}^a\,_c \equiv \delta^a_b- \Pi_{\parallel}^a\,_b
\eeq
where $\Pi_{\perp}^a\,_c$ is the transverse projector. Since $S_{ab}$  is not invertible itself, it is useful to define at this point
\beq
\Box_{ab} \equiv S_{ab} + D^{\ha}_{a} \eta_{\ha\hb} D^{\hb}_{b} 
\eeq
which is invertible with the propagator
\beq
\Box_{ab}p^{bc} = \delta^c_a
\eeq 

Then one can check, using $D^{\ha}_a P^{ab} =0$ and \eq{S2P}, that the propagator $p^{ab}$ is given by 
\beq 
p^{ab} = P^{ab} + T^a_{\a} \eta^{\a\b} T_{\b}^b
\eeq
and thus 
\beq
P^{ab} =  p^{ab}  - T^a_{\a} \eta^{\a\b} T_{\b}^b
\eeq 
Equally $P^{ab}$ satisfies
\beq
  \Box_{ac}  P^{cb} = \Pi_{\perp}^b\,_a
\eeq
So we can write it as 
\beq
 P^{ab} = \Pi_{\perp}^a\,_c p^{cb}\,,
\eeq
or in an explicitly symmetric form as 
\beq
 P^{ab} = \Pi_{\perp}^a\,_c p^{cd} \Pi_{\perp}^b\,_d \,.
\eeq
We also have the nice identities 
\beq
p^{ab} D_{b}^{\a} = - T^a_\b \eta^{\b\a} 
\eeq
and
\beq
\Box_{ab} T^b_\ha = - D_a^\hb \eta_{\hb \ha}
\eeq

\subsection{One-loop}
For our choice of gauge the on-shell effective action is given by 
\beq
e^{-\Gamma} = e^{-S} \frac{\sqrt{\det \gamma_{ab}} }{\sqrt{ \det
\begin{pmatrix}
S_{ab} \, & D^{\hat{\b}}_{a} \\
 D^{\hat{\a}}_{b}\, & 0 
\end{pmatrix} }  \sqrt{\det \eta_{\a\b}}}
\eeq
where we note that the ghosts do not contribute since the FP determinant is one.
Using $\gamma_{ab}$ and $\eta_{\a\b}$ to raise and lower indices it follows that 
\bea
&&
\det
\begin{pmatrix}
S_{ab} \, & D^{\hat{\b}}_{a} \\
 D^{\hat{\a}}_{b}\, & 0 
\end{pmatrix}
\det \gamma^{ab} \det  \eta_{\a\b}
=
\det \begin{pmatrix}
S^a\,_b \, & D^{a}_{\hat{\beta}} \\
 D^{\hat{\a}}_{b}\, & 0 
\end{pmatrix}     \nonumber
\eea
Then we have that
\bea
\det \begin{pmatrix}
S^a\,_b \, & D^{a}_{\hat{\beta}} \\
 D^{\hat{\a}}_{b}\, & 0 
\end{pmatrix}      &=& \det \begin{pmatrix}
S^a\,_b \, & D^{a}_{\hat{\beta}} \\
 D^{\hat{\a}}_{b}\, & 0 
\end{pmatrix} 
\det \begin{pmatrix}
p^a\,_b \, & 0 \\
0\, & \delta^\a_\b 
\end{pmatrix}  
\det \Box^a\,_b\nn
&=&
\det \begin{pmatrix}
\Pi^a_\perp\,_b \, & D^{a}_{\hat{\beta}} \\
 -T^{\hat{\a}}_{b}\, & 0 
\end{pmatrix} 
\det \Box^a\,_b
\eea
Then we note that 
\beq
\det \begin{pmatrix}
\Pi^a_\perp\,_b \, & D^{a}_{\hat{\beta}} \\
 -T^{\hat{\a}}_{b}\, & 0 
\end{pmatrix}  =1
\eeq
since the square of the matrix is the identity
\beq
 \begin{pmatrix}
\Pi^a_\perp\,_b \, & D^{a}_{\hat{\beta}} \\
 -T^{\hat{\a}}_{b}\, & 0 
\end{pmatrix}^2 = 
\begin{pmatrix}
\delta^a\,_b \, &0  \\
0 \, & \delta^\ha_\hb
\end{pmatrix}
\eeq
So we find that 
\beq
e^{-\Gamma} = e^{-S} \frac{ 1  }{ \sqrt{ \det \Box^a\,_b  } } =  e^{-S}  \sqrt{ \det p^a\,_b  } 
\eeq
Using the expression for $p^a\,_b$ we deduce that 
\beq
e^{-\Gamma} = e^{-S} \sqrt{ \frac{\det T^{\b}_{b} T^b_\c   }{  \det\nolimits_\perp S^a\,_b   } }
\eeq
where $\det\nolimits_\perp$ indicates that we take the product of the transverse, i.e. non-zero, eigenvalues and we used that the longitudinal eigenvalues of $p^a\,_b$ are equal to those of $T^{\b}_{b} T^b_\c$.

\section{Consequences of BRST symmetry}
\label{sec:BRST_Consequences}
%
%

Let us now review the consequences of BRST for the effective action $\Gamma[\Phi]$.

\subsection{ Gauge independence}
\label{GauInd}
Let's start by noting that sources to entirely gauge invariant operators may be included in $S[\phi]$. Thus we can think of a general $\mathcal{Z}$ as a generating functional for observables if we choose appropriate sources. We have already shown that $\mathcal{Z}$ is independent of the choice of dressing field using gauge invariance. However, BRST symmetry can be used to show the same result.

If we summarises the BRST transformations \eq{eq:BRST} by 
\beq
\delta_\theta \Phi^A = \theta s^A  
\eeq
then we can define
\bea
s &=& s^A \frac{\delta}{\delta \Phi^A}\nn
& =& - T^a_\a[\phi] C^{\a} \frac{\delta}{\delta \phi^a} + \frac{1}{2}  f^{\a}_{\b\c} C^{\b} C^{\c} \frac{\delta}{ \delta C^{\a}}\nn
&&- B_{\hat{\a}} \frac{\delta}{\delta A_{\hat{\a}}} - \bar{B}_{\hat{\a}} \frac{\delta}{\delta \bar{A}_{\hat{\a}}}
\eea
where for gravity the last term is omitted. 
One can then show that $s$ is nilpotent: $s^2 = 0$.
A term that can be written as $s \Upsilon$ is said to be BRST exact.
Furthermore with
\beq
\tilde{\rho} \equiv \tilde{\mu}\, e^{-\S+S_{\rm id} }
\eeq  
where $S_{\rm id}$ is given by \eq{SidYM} for Yang-Mills and \eq{SidGR} for gravity.
BRST invariance of the theory, including the measure, is summarised by 
\beq
 \frac{\delta}{\delta \Phi^A}  \left( s^A  \tilde{\rho}\right) = 0 \,.
\eeq
Then if $\frac{\partial} {\partial \upzeta} \S =  s \Upsilon$ for a parameter $\upzeta$ we get 
\bea
\frac{\partial}{\partial \upzeta}  \mathcal{Z}  &=&-\int d \Phi  \tilde{\rho}  s^A \frac{\delta}{\delta \Phi^A} \Upsilon \nn
& &=  \int d \Phi  \frac{\delta}{\delta \Phi^A} \left( \tilde{\rho}  s^A \right)  \Upsilon \nn
&=& 0\,.
\eea
Thus a BRST exact term does not effect the functional integral  \cite{Zinn-Justin:1974ggz,Zinn-Justin:1989rgp}.
This generalises our previous notation of gauge independence, relating to the choice of dressing field, to actions with quite general BRST exact terms. Indeed, the actions $\S$ for two different choices of dressing fields differ by BRST exact terms.
Thus BRST exact terms are synonymous with generalised gauge fixing terms.
For Yang-Mills the action we have been using is given by
\beq
\S = S - \frac{1}{2} s ( A_{\hat{\a}}^* D^{\hat{\a}} +   A_{\hat{\a}} \bar{D}^{\hat{\a}}) \,,
\eeq
while for gravity
\beq
\S = S -  s  A_{\hat{\a}} D^{\hat{\a}} \,.
\eeq
Thus the difference between actions for different choices of dressing fields are BRST exact.

\subsection{Master equation}
\label{MasterEq}
Because BRST symmetry is non-linear the BRST symmetry of the effective action $\Gamma[\Phi]$ is not the straightforward generalisation of the BRST symmetry of the $\S$ i.e. beyond tree-level $s \Gamma \neq 0$. Instead BRST symmetry of is realised when we couple source to the BRST transformations.

If we take $W[\mathcal{J}] \to W[\mathcal{J},\Phi^\star]  $ to include sources $\Phi^\star_A$, usually called anti-fields \cite{Batalin:1981jr}, to the BRST transformations $s^A$ 
such that
\beq
\S[\Phi] \to \S[\Phi] -  s^A[\Phi] \Phi^\star_A 
\eeq
then we continue to preserve BRST symmetry since $s^A[\Phi] \Phi^\star_A = s \Phi^A \Phi^\star_A $  is BRST exact.  Note the $B_{\hat{\a}}$-components of $\Phi^\star_A$ do not contribute and so we can put them to zero.
Then we can define now
\beq
\tilde{\rho}[\Phi,\Phi^\star] \equiv \tilde{\mu}\, e^{-\S+s^A[\Phi] \Phi^\star_A  + \dots }\,.
\eeq  
 Since with these sources
\beq
\tilde{\rho} \frac{\overleftarrow{\delta}  }{\delta \Phi^\star_A} = s^A   \tilde{\rho}\,,
\eeq
 the BRST invariance of the theory is summarised by the classical master equation\footnote{ By writing the master equation this way we take into account the measure properly and we assume there is no anomaly.  }
\beq
\frac{ \overrightarrow{\delta}  }{ \delta \Phi^A }  \tilde{\rho}  \frac{\overleftarrow{\delta}  }{\delta \Phi^\star_A}  = 0 \,.
\eeq
Then we have that 
\bea \label{Schwinger}
&&e^{W[\mathcal{J}]} = \int d\Phi \,  \tilde{\rho}[\Phi,\Phi^\star] e^{\mathcal{J}_A  \Phi^A }  
\eea

For $W[\mathcal{J}]$ the master equation implies (by integrating by parts) that 
\beq
\mathcal{J}_A \frac{\delta W }{\delta \Phi^\star_A} = 0
\eeq
and for the effective action we have the master equation \cite{Zinn-Justin:1974ggz,Zinn-Justin:1989rgp}
\beq \label{master}
\Gamma \frac{\overleftarrow{\delta}  }{\delta \Phi^A} \frac{\overrightarrow{\delta} }{\delta \Phi^\star_A} \Gamma  = 0\,.
\eeq
The power of the master equation is that it is stable under renormalisation even if we use a scheme where the BRST terms in our original ansatz is modified by counter terms.
One can introduce a nilpotent operator quantum BRST \cite{Zinn-Justin:1989rgp}
\beq
r = \Gamma  \frac{ \overleftarrow{\delta} }{\delta \Phi^A}\frac{\overrightarrow{\delta}}{ \delta \Phi_A^\star} -  \Gamma  \frac{\overleftarrow{ \delta}     }{\delta \Phi_A^\star}\frac{\overrightarrow{\delta}}{ \delta \Phi^A}
\eeq
and write the master equation as
\beq
r \Gamma =  0\,.
\eeq
 A term in $\Gamma$ which can be written $r \Upsilon$ is BRST exact. 
 
 A useful identity follows  by taking a $\Phi$-derivative of the master equation \eq{master}:

\beq  \label{master1}
-  \Gamma  \frac{\overleftarrow{\delta}   }{\delta \Phi^\star_B}  \,  \frac{\overrightarrow{\delta} }{\delta \Phi^B}   \Gamma \frac{\overleftarrow{\delta} }{ \delta \Phi^A}  +        \Gamma   \frac{\overleftarrow{\delta}  }{\delta \Phi^B} \frac{\overrightarrow{\delta}  }{\delta \Phi_B^\star}   \Gamma  \frac{\overleftarrow{\delta} }{\delta \Phi^A}   = 0\,. 
\eeq
which implies that 
\beq \label{master1b}
 \Gamma  \frac{\overleftarrow{\delta} }{\delta \Phi^\star_C}  =   \Gamma   \frac{\overleftarrow{\delta}  }{\delta \Phi^B} \frac{\overrightarrow{\delta}  }{\delta \Phi_B^\star}   \Gamma  \frac{\overleftarrow{\delta} }{\delta \Phi^A}    P^{AC}\,.
 \eeq

\subsection{Gauge invariance of $\Gamma[\Phi,\Phi^\star]$}

To use the master equation we have to use an effective action $\Gamma[\Phi, \Phi^\star]$ with the anti-fields $\Phi^\star =\{A^{\ha}_\star, \bar{A}^{\ha}_\star,  \phi^\star_a, C^\star_\a \}$. This effective action is also gauge invariant with non-trivial transformation laws of the anti-fields such that
\beq
\Gamma[\g_*\Phi,\g_* \Phi^\star] =  \Gamma[\Phi, \Phi^\star] 
\eeq
This follows from the invariance of $\Phi^\star_A s^A[\Phi]$.  Like the master equation, we expect this symmetry to be preserved if we use a symmetric regulator. 

 For Yang-Mills we have 
\bea
s^A \Phi^\star_A  &=&  \int_x \left(  \nabla_{\mu} C^i  {\rm a}_{\star i}^\mu + \frac{1}{2}  f^k_{i j} C^i C^j C^\star_k  \right) \nn
&&+  \int_x \tr \left(     {\bf B}  {\bf A}_\star^T  +  {\bf \bar{B}} {\bf \bar{A}}_\star^T   \right)
\eea
We can rewrite
\beq
 \nabla_{\mu} C^i  {\rm a}_{\star i}^\mu= \frac{1}{2} \tr \,   \left( \partial_\mu {\bf C} - \im [{\bf  a}_\mu,  {\bf C}] \right) {\bf a}_{\star}^\mu
\eeq
so the transformation law is given by 
\beq
\g_* {\bf a}_{\star}^\mu =  {\bf  G}[\g]  {\bf  a}_\star{\bf G}^{-1}[\g]
\eeq
Also we have
\beq
 \frac{1}{2}  f^k_{i j} C^i C^j   C^\star_k= - \im 2 \tr  \bf{C}^2 {\bf C}^\star_k 
\eeq
from which we infer that
\beq
\g_* {\bf C}_{\star} =  {\bf  G}[\g]  {\bf C}_{\star}{\bf G}^{-1}[\g]\,.
\eeq
Then finally we have that
\beq
\g_*{\bf A}_{\star} = {\bf  A}_{\star}{\bf \bar{G}}^{-1}[\g]\,.
\eeq

For gravity, a similar analysis determines the transformations: $g^{\mu\nu}_\star$ transforms as a contravariant tensor density of weight one,  $C_{\mu}^{\star}$  transforms as a covariant vector density of weight one and $A^{\hat{\mu}}_\star$ transform as four scalars.

\section{Renormalisability of the gauge condition}
\label{sec:Gauge_Condition}
 
Let us address an important point that we have so far neglected. For a general gauge we will encounter UV divergencies proportional to BRST exact terms. In general UV divergencies can  by removed by a renormalisation of the coupling in $\S[\Phi]$ such that $\Gamma[\Phi]$ remains finite in the limit that the cutoff is removed \cite{Gomis:1995jp}. One might then conclude that our choice of gauge is not stable under renormalisation for most choices of dressing fields. This seems to undermine gauge independence since we could not then simply pick our desired gauge.  

As such we would like to address the extent to which we can maintain a choice of gauge after renormalisation.
We will have in mind a gauge invariant regularisation of the theory with a cutoff scale $\Lambda$ that can be achieved through a combination of covariant higher derivatives and Pauli-Villars regularisation schemes \cite{Slavnov:1972sq,Lee:1972fj,Slavnov:1977zf,Faddeev:1980be,Asorey:1989ha,Bakeyev:1996is,Falls:2020tmj}.

In a general non-perturbative scheme we expect all couplings consistent with the symmetries to be renormalised.
 However, certain couplings need not be renormalised if a change in their value can be absorbed by a field redefinition. These couplings are called inessential couplings and do not enter expressions for observables \cite{Wegner:1974sla,Weinberg:1980gg}. Since we have already shown that observables are independent of BRST exact terms we expect that all couplings to   BRST exact terms are inessential. Let us now show this is indeed the case.



To this end let's now rename the ``fundamental fields'' $\mathcal{X}^A$ such in $\S$ and the measure we replace $\Phi$ by $\mathcal{X}$ (i.e. we rename the integration variable).  Then we couple the source to  $ \Phi_\upzeta[\mathcal{X}]$ where $\upzeta$ is by definition an inessential coupling and  $ \Phi_\upzeta[\mathcal{X}]$ field reparameterisation. Thus we obtain the generalised Schwinger functional
\bea \label{Schwinger_zeta}
&&e^{W_\upzeta[\mathcal{J}]} = \int d\X \, \tilde{\mu}[\X] e^{-\S[\X] +  \mathcal{J}_A  \Phi^A_\upzeta[\X] }  
\eea
that depends on the inessential coupling $\upzeta$. The corresponding effective action $\Gamma_\zeta[\Phi]$ will be gauge invariant if 
\beq \label{Cov_Field}
\Phi^A_\upzeta[\g_* \mathcal{X}] =\g_* \Phi^A_\upzeta[\mathcal{X}]\,,
\eeq
which states that $\Phi^A_\upzeta[\mathcal{X}]$ transform covariantly with respect to the original fields $\X^A$.  
 Since the divergencies we need to deal with will also by gauge invariant we expect this to be the case for the required form of  $\Phi^A_\upzeta[\mathcal{X}]$.
 
 One can then show that the effective action depends on $\upzeta$ such that \cite{Wegner:1974sla,Weinberg:1980gg,Tyutin:2000ht,Pawlowski:2005xe,Baldazzi:2021ydj} 
\beq \label{Red_OP}
\frac{\partial}{\partial \upzeta}  \Gamma_\upzeta[\Phi] = -  \frac{\delta \Gamma_\upzeta[\Phi]}{\delta \Phi^A} \Psi^A \,,
\eeq
with
\beq
\Psi_\upzeta = \left\langle  \frac{\partial}{\partial \upzeta}  \Phi_\upzeta[\X'] \right\rangle
\eeq
 A change in $\upzeta$ can therefore be compensated by a field redefinition. One then says that $\upzeta$ is the inessential coupling conjugate to the redundant operator which is the rhs of  \eq{Red_OP}.

Next we note that that every  BRST exact term is a redundant operator i.e. that
\beq \label{RedBRSTexact}
r \Upsilon =-\frac{\delta \Gamma}{\delta \Phi^A}  \Psi^A 
\eeq
for some $\Psi^A$.   To see this we couple the anti-fields to $s \Phi_\zeta[\mathcal{X}]$ such that 
\bea \label{Schwinger_zeta}
&&e^{W_\upzeta[\mathcal{J},\Phi^\star]} = \int d\X \, \tilde{\mu}[\X] e^{-\S[\X] +  \mathcal{J}_A  \Phi^A_\upzeta[\X]+  s \Phi^A_\upzeta[\X]    \Phi_A^\star }  \nonumber
\eea
then $r$ is defined with respect to the corresponding effective action $\Gamma_\upzeta[\Phi,\Phi^\star]$.
We then write $r \Upsilon$ explicitly:
\beq
r \Upsilon =  \Gamma  \frac{ \overleftarrow{\delta} }{\delta \Phi^A}\frac{\overrightarrow{\delta}}{ \delta \Phi_A^\star} \Upsilon -  \Gamma  \frac{\overleftarrow{ \delta}     }{\delta \Phi_A^\star}\frac{\overrightarrow{\delta}}{ \delta \Phi^A}  \Upsilon 
\eeq
noticing that the first term is evidently of the desired redundant form while second term is redundant due to the identity \eq{master1b}.  Now we observe that if $\upalpha$ is a coupling conjugate to a BRST exact term in $\S$, i.e.
\beq 
 \frac{\partial}{\partial \upalpha} \S = s \Upsilon'_\upalpha\,,
\eeq
then
\beq
\frac{\partial}{\partial \upalpha}  \Gamma= \Gamma \frac{\overleftarrow{\delta}  }{\delta \Phi^A} \langle s^A \Upsilon'_\upalpha  \rangle \,\,\,\,  
\eeq
which can also be written as 
\beq
\frac{\partial}{\partial \upalpha}  \Gamma= r  \Upsilon_{\upalpha}  \,\,\,\, {\rm with} \,\,\,\,\, \Upsilon_{\upalpha} =  \langle \Upsilon'_{\upalpha} \rangle\,.
\eeq
Thus we can conclude that $\upalpha$ is an inessential coupling conjugate to BRST exact term in $\Gamma$.
Consequently instead of having to renormalise the BRST exact part of action we can renormalise the field.

In particular let's consider different ways that we can keep $\Gamma$ independent of $\Lambda$. First suppose that for fixed $\Phi[\X]$ it is required that  $\upalpha = \upalpha(\Lambda)$ depends on the UV cutoff $\Lambda$ such that $\Gamma$ is independent of $\Lambda$. Then if we were to fix $\upalpha$ in $\S$ we have that 
\beq
\frac{\partial  \Gamma }{\partial \Lambda} = -\frac{\partial  \upalpha}{\partial \Lambda} \, \Gamma \frac{\overleftarrow{\delta}  }{\delta \Phi^A} \langle s^A \Upsilon'_\upalpha  \rangle
\eeq
But this can then be compensated by letting $\Phi_{\zeta = \upalpha(\Lambda)}[\X]$ depend on $\Lambda$ with 
\beq
\frac{\partial}{\partial \upalpha} \Phi^A_\upalpha[\X] = - s \Phi^A_\upalpha[\X] \Upsilon'_\upalpha[\X]\,.
\eeq
Note that this equation is covariant and thus consistent with \eqref{Cov_Field} provided $\Upsilon'_\upalpha[\X]$ is gauge invariant. As a result  of the above analysis only $S_\Lambda[\phi]$ and $\Phi_\Lambda[\X]$ need to depend on the cutoff $\Lambda$ in order that $\Gamma[\Phi]$ is independent of $\Lambda$. Then at some microscopic scale $\Lambda_0$ we can set boundary conditions such that
\beq
 \Phi^A_{\Lambda_0}[\X] = \X^A
\eeq
In this sense we can pick our desired gauge fixing condition. 

\section{Renormalised gauge invariance}
\label{sec:renGI}
Let us now discuss the case where we remove the UV cutoff $\Lambda = \Lambda_0 \to \infty$. Suppose we have achieved this limit and kept $\Gamma[\Phi]$ finite for some choice of dressing field. Then correlation functions of 
$\hat{\phi}^a$ are finite. If we consider another set of dressed observables we can write these has functions $\hat{\varphi}^{a} = E^{a}[\hat{\phi}]= \mathfrak{E}[\hat{\phi}]_*\hat{\phi}^{a}$ of the original set, as explained in section~\ref{sec:Ex}. However these will involve products of the fields (or its derivatives) at coincidence points and thus the correlation functions of  $\hat{\varphi}^{a}$ will {\it not} be finite in general when the map $E^{a}[\cdot]$ is itself finite.

 What we expect instead is a set of renormalised operators gauge invariant $\hat{\phi}^a_{n,\Lambda}$ (with $\hat{\phi}_{0,\Lambda}^a \equiv \phi^a$) which replace the naive basis of operators. For example, at a Gaussian fixed point the naive operators are replaced by the normal ordered operators.  Then correlation functions of a general  renormalised field  
\beq
\hat{O}^{a} = \sum_{n=0} \hat{O}_n \hat{\phi}^a_{n,\Lambda}[\hat{\phi}]
\eeq   
will have finite correlation functions for finite coefficients $O_n$ even in the limit $\Lambda \to \infty$. For finite cutoff the operators $ \hat{\phi}^a_{n,\Lambda}$ will be a regular basis, but in the limit $\Lambda \to \infty$ they will be singular. 

Thus to understand gauge invariance in the continuum limit we must consider {\it singular} gauge transformations. These can be obtained as follows. We take a general new dressed field $\hat{\varphi}^{a}$ at finite $\Lambda= \Lambda_0$ and expand it  
\beq
\hat{\varphi}^{a} =  \mathfrak{E}[\hat{\phi}]_*\hat{\phi}^{a} =  E^{a}[\hat{\phi}]  = \sum_{n=0} E_n \hat{\phi}^a_{n,\Lambda_0}[\hat{\phi}]
\eeq 
such now the coefficients are constrained $E_n$ such that the second equality holds for some exchanger $\mathfrak{E}[\hat{\phi}]$. Since we are at finite $\Lambda_0$ we can consider a finite exchanger expand it in the renormalised basis and read of the coefficients $E_n$. Then in the continuum limit with $E_n$ fixed we get a singular expression 
\beq
E^a_{\rm div}[\hat{\phi}] = \lim_{\Lambda \to \infty}   \sum_{n=0} E_n \hat{\phi}^a_{n,\Lambda}[\hat{\phi}]
\eeq
Thus in the continuum limit we expect that we need singular gauge transformations that relate dressed fields with finite correlations functions.

\section{Discussion}
\label{sec:Diss}
We have considered in this paper a new fundamental form of the gauge fixed functional integral in gauge theories and quantum gravity. This form follows naturally from simply picking a single representative from each equivalence class of field histories. In turn this leads to a natural corresponding BRST symmetric action. Remarkably, these actions also enjoys  full gauge and diffeomorphism invariance for the respective theories. This clarifies an important point: BRST symmetry does not replace gauge invariance but can live along side it, since here we have both types of invariances enjoyed by one and the same action $\S$. While BRST is a non-linear symmetry, gauge invariance is linear in the fields $\phi^a$ and thus preserving it as a symmetry of the effective action can be a powerful technical advantage. To realise both symmetries requires the ghosts, anti-ghosts and Nakanishi-Lautrup fields.

If one would like to compute correlation functions of dressed fields our effective action allows this computation in a more direct manner than the standard one by choosing the corresponding dressing field. Alternatively one can keep this choice unspecified and compute an observable of choice without ever fixing the explicit choice of gauge.

Under renormalisation the master equations will be satisfied if a symmetric regularisation is used.
The action that obeys the master equation is gauge invariant. This implies we can carry our renormalisation without breaking gauge invariance in its extended form. One may hope that the enhanced symmetry gives new insights into gauge theories and quantum gravity.

 In section~\ref{sec:Gauge_Condition} we have argued that BRST exact terms are inessential. Therefore one is not forced to renormalise the gauge.  However, the conclusion drawn in section~\ref{sec:renGI} implies that the relation between effective actions, which differ by the choice of dressing field, will in fact involve singular transformations in the continuum limit.

\section*{Acknowledgements}
It is a pleasure to thank Renata Ferrero, Boris Stupovski and Antonio Pereira  for many discussions and for their comments on the paper. I also thank Kristina Giesel and Thomas Thiemann for their useful comments and questions during a presentation of this work at Friedrich-Alexander University.

\begin{appendix}

\section{Yang-Mills theory}
\label{App_YM}
Here we collect results that concern Yang-Mills theory. 
Let's define the n-fold commutator by 
\beq
[{\bf M},{\bf N}]_n =  [[{\bf M},{\bf N}]_{n-1}, {\bf N}]\,,
\eeq
with $[{\bf M},{\bf N}]_0 = {\bf M}$. Then
\beq
e^{\bf N} {\bf M } e^{-{\bf N}} = \sum_{n=0}^\infty (-1)^n \frac{1}{n!} [{\bf M},{\bf N}]_n \,.
\eeq
Then we note that by induction we can prove that 
\beq
[{\bf t}_i, {\rm i}  \g^j {\bf t}_j]_n = (-1)^n (F^n(\g))^{ik} {\bf t}_{k}
\eeq
where $F$ is the matrix with components  $(F(\g))^{ik} = f^{ijk}\g^j$ and $F^n$ is the nth power of this matrix.
Hence we obtain 
\bea
e^{ {\rm i}  \g^j {\bf t}_j}  {\bf t}_i  e^{- {\rm i}  \g^j {\bf t}_j} &=&  \sum_{n=0}^\infty  \frac{1}{n!}  ( F^n(\g))^{ik} {\bf t}_{k} \nn
& =&  \left(e^F(\g)\right)^{ik}  {\bf t}_{k}\nn
& =&    {\bf t}_{k} \left(e^{-F(\g)}\right)^{ki}
\eea
this is a useful identity which is \eq{amarillo} in the main text. We will also use it in the following form
\beq
e^{-{\rm i} \g_j {\bf t}_j}  {\bf t}_i =  {\bf t_k}    e^{-{\rm i} \g_j {\bf t}_j}   \left(e^{F(\g)}\right)^{ki} 
\eeq

Then we note that since $\mathfrak{1}^i = 0$ we have
\beq
{\rm i} \left. \frac{\partial}{\partial \h^i} (\g_*\h_*\hat{\g})^j {\bf t}_j  \right|_{\h = \mathfrak{1}} =  \left. \frac{\partial}{\partial \h^i}  e^{ {\rm i}  \g_*\h_*\hat{\g}^j {\bf t}_j}    \right|_{\h = \mathfrak{1}}   
\eeq
which on the other hand is equal to 
\bea
{\rm i}  \left. \frac{\partial}{\partial \h^i} (\g_*\h_*\hat{\g})^j {\bf t}_j  \right|_{\h = \mathfrak{1}} &=&    \left. \frac{\partial}{\partial \h^i}  e^{ {\rm i}  \g^j {\bf t}_j}  e^{ {\rm i}  \h^j {\bf t}_j}  e^{ - {\rm i}  \g^j {\bf t}_j}    \right|_{\h = \mathfrak{1}} \\
  &=&  e^{ {\rm i}  \g^j {\bf t}_j} {\rm i} {\bf t_i } e^{- {\rm i}  \g^j {\bf t}_j} 
\eea
Thus we obtain 
\beq
\left. \frac{\partial}{\partial \h^i} (\g_*\h_*\hat{\g})^k   \right|_{\h = \mathfrak{1}}  = \left(e^F\right)^{ik}  =  \left(e^{-F}\right)^{ki}
\eeq
which demonstrates \eq{Talphabeta_YM}.

\subsection{Transform of $\A^i_\mu$}
\label{App:TransA}
Here we show how the gauge field transforms. From \eq{boldA_transform} and using \eq{amarillo} we have that
\bea
\g_*\A^i_\mu {\bf t}_i &=& e^{{\rm i} \g^j {\bf t}_j} \A^i_\mu {\bf t}_i  e^{-{\rm i} \g^j {\bf t}_j}  +   e^{{\rm i} \g^j {\bf t}_j} \partial_\mu \g^i {\bf t}_i   e^{-{\rm i} \g^j {\bf t}_j}\nn
&=& \A^i_\mu  \left(e^F\right)^{ik}  {\bf t}_{k}  + \partial_\mu \g^i \left(e^F\right)^{ik}  {\bf t}_{k} 
\eea
So we read off that
\beq
\g_*\A^k_\mu =   \A^i_\mu  \left(e^F\right)^{ik}  + \partial_\mu \g^i \left(e^F\right)^{ik} 
\eeq
this can be written as
\beq
\g_*\A^k_\mu = \left(e^{-F}\right)^{ki}  \A^i_\mu   +  \left(e^{-F}\right)^{ki}  \partial_\mu \g^i 
\eeq
Then if we expand to linear order we find that 
\beq
T^a_\a[\phi] = T^i_{\mu j} \eps^j = \partial_\mu \eps^i + f^{ijk} A^j_\mu \eps^k
\eeq
where $T^a_\a[\phi] $ is defined in \eq{Tphi_def}.
\\

\section{Transformation of the measure}
\label{Measure_Trans}

It is a straightforward exercise to prove that 
\beq
\det \delta^a_b  -  \frac{\delta \hat{\mathfrak{E}}^\a}{\delta \phi^b} U_\a\,^\b[\mathfrak{E}] T_\b^a[\phi]  = \det \delta^\a_\b  -  T_\b^a[\phi]  \frac{\delta \hat{\mathfrak{E}}^\c}{\delta \phi^a} U_\c\,^\a[\mathfrak{E}] \,.
\eeq
To do so one proves that for every eigenvector of the operators, that do not have an eigenvalue equal to one, there is a corresponding eigenvector with the same eigenvalue for the other operator. 
Then let's note that
\beq
T_\b^a[\phi]  \frac{\delta \hat{\mathfrak{E}}^\c}{\delta \phi^a} = \left. \frac{\delta}{\delta \g^\b}  \hat{\mathfrak{E}}^\c[\g_*\phi] \right|_{\g= \mathfrak{1}}
\eeq
and 
 \beq
  U_\c\,^\a[\mathfrak{E}]  =   \left. \frac{\delta  \h_*\mathfrak{E}^\a }{\delta  \h^\c}\right|_{\h = \hat{\mathfrak{E}} }
 \eeq
and hence
\beq
T_\b^a[\phi]  \frac{\delta \hat{\mathfrak{E}}^\c}{\delta \phi^a} U_\c\,^\a[\mathfrak{E}]  =   \left. \frac{\delta}{\delta \g^\b}  \hat{\mathfrak{E}}^\c[\g_*\phi] \right|_{\g= \mathfrak{1}}   \left. \frac{\delta  \h_*\mathfrak{E}^\a }{\delta  \h^\c}\right|_{\h = \hat{\mathfrak{E}} }
\eeq
by use of the chain rule this is equal to
\beq
T_\b^a[\phi]  \frac{\delta \hat{\mathfrak{E}}^\c}{\delta \phi^a} U_\c\,^\a[\mathfrak{E}] = \left.  \frac{\delta  \hat{\mathfrak{E}}[\g_*\phi]_*\I^\a[\phi] }{ \delta \g^\b} \right|_{\g= \mathfrak{1}}
\eeq
then using the transformation law for an exchanger we have
\beq
T_\b^a[\phi]  \frac{\delta \hat{\mathfrak{E}}}{\delta \phi^a} U_\c\,^\a[\mathfrak{E}] = \left.  \frac{\delta \g_*\hat{\mathfrak{E}}^\c[\phi]_*\hat{\g}_*\I^\a[\phi]}{ \delta \g^\b} \right|_{\g= \mathfrak{1}}\,.
\eeq
 Expanding we get
\bea
\left.  \frac{\delta \g_*\hat{\mathfrak{E}}[\phi]_*\hat{\g}_*\I^\a[\phi]}{ \delta \g^\b} \right|_{\g= \mathfrak{1}} &=& \delta^\a_\b +  \left.  \frac{\delta \hat{\mathfrak{E}}[\phi]_*\hat{\g}_*\I^\a[\phi]}{ \delta \g^\b} \right|_{\g= \mathfrak{1}}\nn
&=&  \delta^\a_\b -  \left.  \frac{\delta \hat{\mathfrak{E}}[\phi]_*\g_*\I^\a[\phi]}{ \delta \g^\b} \right|_{\g= \mathfrak{1}} \nonumber\,,
\eea
and thus we obtain \eq{Rosa}.

\end{appendix}

\end{document}